\documentclass[prc,preprint,superscriptaddress,showpacs,nofootinbib,tightenlines,amsmath]{revtex4}
\usepackage{graphics}
\usepackage{epsfig}
\usepackage{slashed}
\def\bra#1{\ensuremath{\langle{#1}\vert}}
\def\ket#1{\ensuremath{\vert{#1}\rangle}}
\begin{document}
\title{Pion photo- and electroproduction with chiral MAID}
\author{M.~Hilt}
\affiliation{PRISMA Cluster of Excellence, Institut f\"ur Kernphysik, Johannes Gutenberg-Universit\"at Mainz,
D-55099 Mainz, Germany}
\author{B.~C.~Lehnhart}
\affiliation{PRISMA Cluster of Excellence, Institut f\"ur Kernphysik, Johannes Gutenberg-Universit\"at Mainz,
D-55099 Mainz, Germany}
\author{S.~Scherer}
\affiliation{PRISMA Cluster of Excellence, Institut f\"ur Kernphysik, Johannes Gutenberg-Universit\"at Mainz,
D-55099 Mainz, Germany}
\author{L.~Tiator}
\affiliation{PRISMA Cluster of Excellence, Institut f\"ur Kernphysik, Johannes Gutenberg-Universit\"at Mainz,
D-55099 Mainz, Germany}

\date{September 13, 2013}
\preprint{MITP/13-053}
\begin{abstract}
   We present a calculation of pion photo- and electroproduction in
manifestly Lorentz-invariant baryon chiral perturbation theory
up to and including order $q^4$.
   We fix the low-energy constants by fitting experimental data in all available reaction channels.
   Our results can be accessed via a web interface, the so-called chiral MAID.
   We explain how our program works and how it can be used for further analysis.
\end{abstract}

\maketitle

\section{Introduction}
   The pion triplet ($\pi^+,\pi^0,\pi^-$) comprises the lightest hadrons which are of fundamental
importance in our understanding of the strong interactions.
   In 1935, Yukawa introduced a field mediating the interaction between the proton and
the neutron to explain the nature of the forces in the nucleus \cite{Yukawa:1935xg}.
   Based on experimental results for the mass defect of deuterium,
he estimated the mass associated with the quantum field of the exchanged particle to
be $200$ times as large as the electron mass.
   From the present-day perspective, one-pion exchange is responsible for the long-range part
of the nucleon-nucleon interaction (see, e.g., Refs.~\cite{Epelbaum:2008ga,Machleidt:2011zz} for a
review).
   In 1947, charged pions, produced by cosmic rays at high altitudes, were discovered by Lattes
et al.\ \cite{Lattes:1947mw} in terms of their decay into muons and neutrinos.
   Subsequently, charged pions were produced in the laboratory by impinging alpha particles on a carbon
target \cite{Gardner:1948,Burfening:1987na}.
   On the other hand, neutral pions were first produced in terms of proton-nucleon collisions in nuclei
\cite{Bjorklund:1950zz} and photoproduction on nuclei \cite{Panofsky:1950gj}.

   Ever since the nineteen-fifties, the electromagnetic production of pions on the nucleon has
been an important source of information on the pion-nucleon interaction.
   On the theoretical side, the low-energy theorem (LET) of Kroll and Ruderman \cite{Kroll:1953vq} provided a
prediction for the matrix element for charged pion photoproduction at threshold.
   Based on a few assumptions such as covariance, gauge invariance, and renormalizability, the theorem states that
the photoproduction of charged pions at threshold computed to lowest order in the pion-nucleon mass ratio,
$\mu=M_\pi/m_N$, but to arbitrary order in the pion-nucleon coupling constant is equivalent to a
calculation in second-order perturbation theory with pseudoscalar coupling, provided that
the pion-nucleon coupling constant and the nucleon mass are replaced by their renormalized
values.
   It was also shown that the $\pi^0$ production amplitude vanishes in the limit $\mu\to 0$.
   Multipole expansions for photo- and electroproduction were derived in Refs.\ \cite{Chew:1957tf}
and \cite{Dennery:1961zz}, respectively.
   Because of the large value of the pion-nucleon coupling constant, perturbative methods
turned out to be of limited use and, thus, the treatment of pion production focussed on
dispersive techniques (see Refs.~\cite{Hanstein:1996bd,Hanstein:1997tp,Kamalov:2002wk,Pasquini:2004nq,Pasquini:2006yi,
Pasquini:2007fw,Drechsel:2007gz}
for more recent applications).

   A new twist originated from the interpretation of pions as the (almost) massless
Goldstone bosons of a spontaneous breakdown of chiral symmetry
\cite{Nambu:1960xd,Nambu:1961tp,Goldstone:1961eq,Goldstone:1962es}.
   As was first discussed in Ref.~\cite{Nambu:1997wa}, chirality conservation in the strong interactions
results in the bremsstrahlung of soft pions in any reaction with a change of nucleon helicity.
   The consequences of this observation for the case of pion electroproduction were first worked out
by Nambu and Shrauner \cite{Nambu:1997wb}.
   In particular, as a generalization of the Kroll-Ruderman theorem for a virtual photon,
their result for the production of charged pions involved the normalized isovector axial form factor.

   In quantum chromodynamics (QCD), chiral symmetry originates from the zero-mass limit of
$u$ and $d$ quarks in the two-flavor case, with a straightforward generalization if the
strange-quark mass is also taken to zero.
   Although in the pre-QCD era the dynamical origin of chiral symmetry was not known,
the symmetry structure was inferred from electromagnetic and weak hadron currents
and summarized in terms of the so-called current algebra, i.e., equal-time commutation
relations involving vector- and axial-vector currents  (see, e.g.,
Refs.\ \cite{Adler:1968,Treiman:1972,Alfaro:1973}).
   In particular, as first pointed out by Gell-Mann, the equal-time commutation relations
still play an important role even if the symmetry is explicitly broken \cite{GellMann:1962xb}.
   The so-called partially conserved axial-vector current (PCAC) hypothesis
\cite{Nambu:1960xd,Bernstein:1960a,GellMann:1960np,Bernstein:1960b}
assumed that
the divergence of the axial-vector current is proportional to a renormalized pion field
and would disappear in the limit of massless pions.
   Numerous predictions have been derived from current algebra
(see Refs.\ \cite{Adler:1968,Treiman:1972,Alfaro:1973} for an overview).
   For example, as an application to pion photoproduction, Fubini, Furlan, and Rossetti
derived dispersion relations, connecting the isoscalar and isovector anomalous magnetic moments
of the nucleon with the forward production amplitude for soft pions
\cite{Fubini:1965,Pasquini:2004nq,Bernard:2005dj}.
   Another example is given by the Adler-Gilman relation \cite{Adler:1966gd}, providing a
consistency relation for pion electroproduction in terms of a chiral Ward identity
\cite{Adler:1966gd,Fuchs:2003vw}.
   Finally, by including the PCAC hypothesis, corrections for the threshold amplitudes
beyond the LET of Kroll and Ruderman were
investigated in, e.g., Refs.~\cite{DeBaenst:1971hp,Vainshtein:1972ih,Scherer:1991cy}.
   A comprehensive overview of the various phenomenological implications of PCAC and current
algebra for pion electroproduction can be found in Ref.~\cite{Amaldi:1979vh}.

   As an alternative method to the often unwieldy soft-pion techniques, Weinberg
constructed an effective Lagrangian for soft-pion interactions reproducing the results of
current algebra \cite{Weinberg:1966fm}.
   While in the beginning phenomenological Lagrangians were applied with the understanding that
they should only be used at tree level \cite{Weinberg:1966fm,Schwinger:1967tc,Weinberg:1968de,
Coleman:1969sm,Callan:1969sn}, in 1979 it was pointed out by Weinberg \cite{Weinberg:1978kz}
that corrections to the chiral limit could be calculated systematically in terms
of an effective field theory (EFT) program.
   The approach is based on a perturbative calculation using a momentum expansion based on
the most general Lagrangian consistent with chiral symmetry.
   With QCD as the underlying fundamental theory, the corresponding low-energy EFT in terms
of pions and nucleons as effective degrees of freedom is
chiral perturbation theory (ChPT) \cite{Weinberg:1978kz,Gasser:1983yg,Gasser:1987rb}
(see, e.g., Refs.~\cite{Ecker:1994gg,Bernard:1995dp,Scherer:2002tk,Scherer:2012zzd} for an introduction).
   Assigning a suitable order to the explicit symmetry breaking due to the quark masses,
it is possible to include quark-mass effects perturbatively.

   Until the 1980s, there was little doubt concerning the validity of the
low-energy predictions for pion photoproduction.
   In particular, the results for the charged channels, which are dominated by the
Kroll-Ruderman theorem, were in good agreement with the available data \cite{Adamovich:1976}.
   However, renewed interest in neutral pion photoproduction at threshold was
triggered by experimental data \cite{Mazzucato:1986dz,Beck:1990da} which indicated
a serious disagreement with the predictions for the $s$-wave electric dipole
amplitude $E_{0+}$ based on current algebra and PCAC \cite{DeBaenst:1971hp}.
   This discrepancy was explained with the aid of ChPT \cite{Bernard:1991rt}.
   Pion loops, which are beyond the current-algebra framework, generate infrared
singularities in the scattering amplitude which then modify the predicted low-energy
expansion of $E_{0+}$ (see also Ref.\ \cite{Davidson:1993et}).
   Subsequently, several experiments investigated pion photo- and electroproduction in
the threshold region \cite{Welch:1992ex,Wang:1992,Liu:1994,vandenBrink:1995uka,Blomqvist:1996tx,Fuchs:1996ja,
Bergstrom:1996fq,Bernstein:1996vf,Bergstrom:1997jc,Kovash:1997tj,Bergstrom:1998ec,Distler:1998ae,
Liesenfeld:1999mv,Korkmaz:1999sg,Schmidt:2001vg,Merkel:2001qg,baumann,Weis:2007kf,Merkel:2009zz,
Merkel:2011cf,Hornidge:2012ca,Hornidge:2013qka,Lindgren:2013eta}.
   From the theoretical side, all of the different reaction channels of pion photo- and electroproduction
near threshold were extensively investigated by Bernard {\it et al.}~within the framework of heavy-baryon
chiral perturbation theory (HBChPT)
\cite{Bernard:1992qa,Bernard:1992nc,Bernard:1992ys,Bernard:1993bq,Bernard:1994dt,Bernard:1994gm,Bernard:1995cj,
Bernard:1996ti,Bernard:1996bi,Fearing:2000uy,Bernard:2001gz}.
   In the beginning, the manifestly Lorentz-invariant or relativistic formulation of ChPT (RChPT) was
abandoned, as it seemingly had a problem with respect to power counting when loops containing internal nucleon
lines come into play.
   Therefore, HBChPT became a standard tool for the analysis of pion photo- and electroproduction in the
threshold region (see, e.g., Ref.~\cite{FernandezRamirez:2012nw}).
   In the meantime, the development of the infrared regularization (IR) scheme \cite{Becher:1999he}
and the extended on-mass-shell (EOMS) scheme \cite{Gegelia:1999gf,Fuchs:2003qc} offered a solution to
the power-counting problem, and RChPT became popular again.
   For example, pion-nucleon scattering was analyzed at $O(q^4)$ in both IR and EOMS schemes
in Refs.~\cite{Becher:2001hv} and \cite{Chen:2012nx}, respectively, and at $O(q^3)$ in the EOMS
scheme including the $\Delta$ resonance \cite{Alarcon:2012kn}.

   The aim of the present article is twofold.
   First, by presenting a full $O(q^4)$ calculation of
pion photo- and electroproduction in the framework of RChPT,
we extend the results of Ref.~\cite{Hilt:2013uf} for neutral pion photoproduction on the proton.
   Second, we present the so-called chiral MAID ($\chi$MAID) \cite{website}.
   This program, accessible via a web interface,  provides the numerical results of these calculations.

\section{Pion photo- and electroproduction}
\begin{figure}[htbp]
    \centering
        \includegraphics[width=0.4\textwidth]{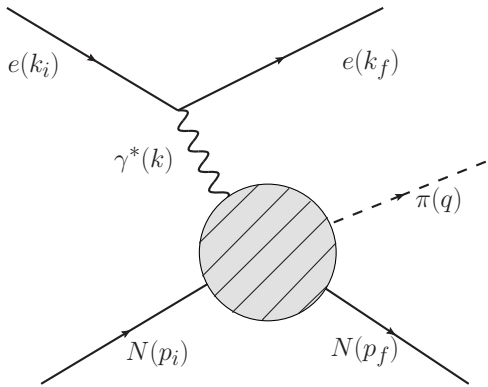}
    \caption{Pion electroproduction in the one-photon-exchange approximation.
    The momenta of the incoming and outgoing nucleons are $p_i$ and $p_f$, respectively.
    The momentum of the incoming/outgoing electron is $k_i$/$k_f$, where $k=k_i-k_f$ represents the
    momentum of the single exchanged virtual photon.
    The momentum of the pion is labeled $q$.
    In the case of pion photoproduction, the leptonic vertex and the photon propagator are
    replaced by the polarization vector of the real photon.
    The shaded circle represents the full hadronic vertex.}
    \label{fig:pionelectro}
\end{figure}

   In this section we provide a short introduction to our notation for describing the
electroproduction of pions,
\begin{equation}
e(k_i)+N(p_i)\rightarrow e(k_f)+N(p_f)+\pi(q).
\end{equation}
   The interaction of the electron with the nucleon is of purely electromagnetic type and, due to the small
coupling $\alpha=e^2/(4\pi)\approx1/137$, the process can be described in the so-called
one-photon-exchange approximation (see Fig.~\ref{fig:pionelectro}).
   In this approximation, the invariant amplitude ${\cal M}$ may be interpreted as the inner product of
the polarization vector $\epsilon_\mu$ of the virtual photon (four-momentum $k=k_i-k_f$) and the
hadronic transition current matrix element ${\cal M}^\mu$,
\begin{equation}
{\cal M}=\epsilon_\mu {\cal M}^\mu,
\end{equation}
where\footnote{For notational convenience, the spin vectors of the electrons and the nucleons are suppressed.
Moreover, we use $e>0$.}
\begin{equation}
\epsilon_\mu=e\frac{\bar{u}(k_f)\gamma_\mu u(k_i)}{k^2}
\label{eq:polarizationvector}
\end{equation}
and
\begin{equation}
{\cal M}^\mu=-ie\langle N(p_f),\pi(q)|J^\mu(0)|N(p_i)\rangle.
\label{eq:mmu}
\end{equation}
   Therefore, it is sufficient to consider the process
\begin{equation}´
\gamma^*(k)+N(p_i)\rightarrow N(p_f)+\pi(q),
\end{equation}
where $\gamma^\ast$ refers to a virtual photon.

   The invariant amplitude of pion photoproduction is obtained by replacing the polarization
vector of the virtual photon by the polarization vector of a real photon and taking $k^2=0$.
   Treating the virtual photon as a particle of ``mass'' $k^2=-Q^2$, the Mandelstam variables
$s$, $t$, and $u$ are defined as
\begin{equation}
s=(p_i+k)^2=(p_f+q)^2,\ u=(p_i-q)^2=(p_f-k)^2, \ t=(p_i-p_f)^2=(q-k)^2,\label{eqn:mandelstam}
\end{equation}
and fulfill
\begin{equation}
s+t+u=2m_N^2+M_\pi^2-Q^2,
\end{equation}
where $m_N$  and $M_\pi$ denote the nucleon mass and the pion mass, respectively.
   In the case of photoproduction ($k^2=0$) only two of the Mandelstam variables are independent.
   In the center-of-mass (cm) frame, the energies of the photon, $k_0$, and the pion, $E_\pi$,
are given by
\begin{equation}
k_0=\frac{W^2-m_N^2-Q^2}{2W},\ E_\pi=\frac{W^2+M_\pi^2-m_N^2}{2W},
\end{equation}
where $W=\sqrt{s}$ is the cm total energy.
   The equivalent real photon laboratory energy $E_\gamma^\textnormal{lab}$ is given by
\begin{equation}
E_\gamma^\textnormal{lab}=\frac{W^2-m_N^2}{2m_N}.
\end{equation}
   The cm scattering angle $\Theta_\pi$ between the pion three-momentum and the $z$-axis,
defined by the incoming (virtual) photon, can be related to the Mandelstam variable $t$ via
\begin{equation}
t=M_\pi^2-2(E_\gamma E_\pi-|\vec{k}||\vec{q}\hspace{0.05cm}|\cos\Theta_\pi).
\end{equation}
   The matrix element of pion electroproduction can be parametrized in terms of the so-called Ball amplitudes
\cite{Ball:1961zza}, which are defined in a Lorentz-covariant way and are convenient for calculating the process
$\gamma^*(k)+N(p_i)\rightarrow N(p_f)+\pi(q)$,
\begin{equation}
-ie\bra{N'\pi}J^\mu(0)\ket{N}
=\bar{u}(p_f)\left(\sum_{i=1}^8 B_i V_i^\mu\right) u(p_i).
\label{eqn:ball}
\end{equation}
   In Eq.\ (\ref{eqn:ball}), $J^\mu$ is the electromagnetic current
operator in units of the elementary charge $e>0$,
and $u(p_i)$ and $\bar{u}(p_f)$ are the Dirac spinors of the nucleon in the initial and
final states, respectively.
   In the following, our convention differs slightly from Ball's original definition.
   We use
\begin{equation}
\begin{split}
V_1^\mu&=\gamma^\mu\gamma_5,\quad
V_2^\mu=\gamma_5 P^\mu,\quad
V_3^\mu=\gamma_5 q^\mu,\quad
V_4^\mu=\gamma_5 k^\mu,\\
V_5^\mu&=\gamma^\mu \slashed{k}\gamma_5,\quad
V_6^\mu=\slashed{k}\gamma_5 P^\mu,\quad
V_7^\mu=\slashed{k}\gamma_5 q^\mu,\quad
V_8^\mu=\slashed{k}\gamma_5 k^\mu,
\end{split}
\label{eq:Ball_structures}
\end{equation}
with $P=(p_i+p_f)/2$ and $\slashed k=\gamma^\mu k_\mu$.
   Electromagnetic current conservation, $k_\mu{\cal M}^\mu=0$, leads to the following two constraints
for the amplitudes $B_i$,
\begin{equation}
\begin{split}
B_1+B_6 k\cdot P+B_7 k\cdot q+B_8k^2&=0,\\
B_2 k\cdot P+B_3 k\cdot q+B_4 k^2+B_5 k^2&=0.
\end{split}
\label{eqn:stromerhalt2}
\end{equation}
   Thus, only six independent amplitudes are required for the description of pion electroproduction.
   Furthermore, in pion photoproduction ($k^2=0$) only four independent amplitudes survive.

   Besides Eq.~(\ref{eqn:ball}), several other parameterizations exist for the matrix element of
Eq.~(\ref{eq:mmu}).
   Here, we focus on those we used for our calculations.
   The parameterization of Ref.~\cite{Drechsel:1992pn} takes care of current conservation already
from the beginning and, thus, contains only six independent amplitudes $A_i$,
\begin{equation}
\mathcal{M}^\mu=\bar{u}(p_f)\left(\sum_{i=1}^6 A_i M_i^\mu\right)u(p_i),
\label{eqn:dennery}
\end{equation}
with
\begin{equation}
\begin{split}
M_1^\mu&=-\frac{i}{2}\gamma_5\left(\gamma^\mu \slashed{k}-\slashed{k}\gamma^\mu\right),\\
M_2^\mu&=2i\gamma_5\left[P^\mu k\cdot\left(q-\frac{1}{2}k\right)
-\left(q^\mu-\frac{1}{2}k^\mu\right) k\cdot P\right],\\
M_3^\mu&=-i\gamma_5\left(\gamma^\mu k\cdot q-\slashed{k}q^\mu\right),\\
M_4^\mu&=-2i\gamma_5\left(\gamma^\mu k\cdot P-\slashed{k}P^\mu\right)-2m_N M_1^\mu,\\
M_5^\mu&=i\gamma_5\left(k^\mu k\cdot q-q^\mu k^2 \right),\\
M_6^\mu&=-i\gamma_5\left(\slashed{k} k^\mu-\gamma^\mu k^2\right).
\end{split}
\end{equation}
   Note that each structure $M_i^\mu$ satisfies $k_\mu M^\mu_i=0$.

   The so-called Chew-Goldberger-Low-Nambu (CGLN) amplitudes $\mathcal{F}_i$ are another
common parameterization \cite{Chew:1957tf,Dennery:1961zz}.
   These amplitudes are defined in the cm frame via
\begin{equation}
\epsilon_\mu\bar{u}(p_f)\left(\sum_{i=1}^6A_i M_i^\mu\right) u(p_i)=\frac{4\pi W}{m_N}\chi_f^\dagger\mathcal{F}\chi_i,
\end{equation}
where $\chi_i$ and $\chi_f$ denote initial and final Pauli spinors.
   Electromagnetic current conservation allows one to work in a gauge where the polarization vector
of the virtual photon has a vanishing time component.
   In terms of the polarization vector of Eq.~(\ref{eq:polarizationvector}) this is achieved by introducing
the vector \cite{Amaldi:1979vh}
\begin{equation}
a^\mu=\epsilon^\mu-k^\mu \frac{\epsilon_0}{k_0}=\epsilon^\mu-k^\mu \frac{\vec{k}\cdot\vec{\epsilon}}{k_0^2},
\end{equation}
where use of $k_\mu \epsilon^\mu=0$ has been made.
   Splitting $\vec a$ into a longitudinal and a transversal piece,
\begin{equation}
\begin{split}
\vec a&=\vec a_\parallel+\vec a_\perp,\\
\vec a_\parallel&=\vec{a}\cdot\hat k\,\hat k=\frac{k^2}{k_0^2}\vec\epsilon\cdot\hat k\,\hat k,\\
\vec a_\perp&=\vec a-\vec a_\parallel=\vec \epsilon-\vec\epsilon\cdot\hat k\,\hat k=\vec\epsilon_\perp,
\end{split}
\end{equation}
$\mathcal{F}$ may be written as
\begin{equation}
\begin{split}
\mathcal{F}=&i \vec{\sigma}\cdot\vec{a}_\perp\mathcal{F}_1
+\vec{\sigma}\cdot\hat{q}\,\vec{\sigma}\cdot\hat{k}\times\vec{a}_\perp\mathcal{F}_2
+i\vec{\sigma}\cdot\hat{k}\,\hat{q}\cdot\vec{a}_\perp\mathcal{F}_3
+i \vec{\sigma}\cdot\hat{q}\,\hat{q}\cdot\vec{a}_\perp\mathcal{F}_4\\
&+i \vec{\sigma}\cdot\hat{k}\,\hat{k}\cdot\vec{a}_\parallel\mathcal{F}_5
+i\vec{\sigma}\cdot\hat{q}\,\hat{k}\cdot\vec{a}_\parallel\mathcal{F}_6,
\end{split}
\label{eq:F}
\end{equation}
where $\hat{q}$ and $\hat{k}$ denote unit vectors in the direction of the pion and the
photon, respectively.
   For the case of pion photoproduction, only the first four terms of Eq.~(\ref{eq:F})
contribute.

   The connection between the Ball amplitudes, invariant amplitudes, and CGLN amplitudes can be
found in Appendix \ref{ampcon}.
   The CGLN amplitudes can be expanded in a multipole series \cite{Chew:1957tf,Dennery:1961zz,Amaldi:1979vh},
\begin{equation}
\begin{split}
\mathcal{F}_1&=\sum_{l=0}^\infty\Big\{\big[lM_{l+}+E_{l+}\big]P'_{l+1}(x)
+\big[(l+1)M_{l-}+E_{l-}\big]P'_{l-1}(x)\Big\},\\
\mathcal{F}_2&=\sum_{l=1}^\infty\Big\{(l+1)M_{l+}+lM_{l-}\Big\}P'_l(x),\\
\mathcal{F}_3&=\sum_{l=1}^\infty\Big\{\big[E_{l+}-M_{l+}\big]P''_{l+1}(x)
+\big[E_{l-}+M_{l-}\big]P''_{l-1}(x)\Big\},\\
\mathcal{F}_4&=\sum_{l=2}^\infty\Big\{M_{l+}-E_{l+}-M_{l-}-E_{l-}\Big\}P''_l(x),\\
\mathcal{F}_5&=\sum_{l=0}^\infty\Big\{(l+1)L_{l+}P'_{l+1}-lL_{l-}P'_l(x)\Big\},\\
\mathcal{F}_6&=\sum_{l=1}^\infty\Big\{lL_{l-}-(l+1)L_{l+}\Big\}P'_l(x),
\end{split}
\label{eq:Fi}
\end{equation}
where $x=\cos\Theta_\pi=\hat{q}\cdot\hat{k}$.
   In Eq.~(\ref{eq:Fi}), $P_l(x)$ is a Legendre polynomial of degree $l$,
$P'_l=dP_l/dx$ and so on, with $l$ denoting the orbital angular momentum
of the pion-nucleon system in the final state.
   The multipoles $E_{l\pm}$, $M_{l\pm}$, and $L_{l\pm}$ are functions of the
cm total energy $W$ and the photon virtuality $Q^2$ and
refer to transversal electric and
magnetic transitions and longitudinal transitions, respectively.
   The subscript $l\pm$ denotes the total angular momentum $j=l\pm1/2$ in the final state.
   By inverting the above equations, the angular dependence can be completely projected out
\cite{Adler:1968tw,Davidson:1995jm},
\begin{equation}
\begin{split}
E_{l+}=&\int_{-1}^1\frac{dx}{2(l+1)}\Big[P_l \mathcal{F}_1-P_{l+1}\mathcal{F}_2\\
&+\frac{l}{2l+1}(P_{l-1}-P_{l+1})\mathcal{F}_3+\frac{l+1}{2l+3}(P_l-P_{l+2})\mathcal{F}_4\Big],\\
E_{l-}=&\int_{-1}^1\frac{dx}{2l}\Big[P_l\mathcal{F}_1-P_{l-1}\mathcal{F}_2\\
&-\frac{l+1}{2l+1}(P_{l-1}-P_{l+1})\mathcal{F}_3+\frac{l}{2l-1}(P_l-P_{l-2})\mathcal{F}_4\Big],\\
M_{l+}=&\int_{-1}^1\frac{dx}{2(l+1)}\left[P_l\mathcal{F}_1-P_{l+1}\mathcal{F}_2
-\frac{1}{2l+1}(P_{l-1}-P_{l+1})\mathcal{F}_3\right],\\
M_{l-}=&\int_{-1}^1\frac{dx}{2l}\left[-P_l\mathcal{F}_1+P_{l-1}\mathcal{F}_2
+\frac{1}{2l+1}(P_{l-1}-P_{l+1})\mathcal{F}_3\right],\\
L_{l+}=&\int_{-1}^1\frac{dx}{2(l+1)}\left[P_{l+1}\mathcal{F}_6+P_l\mathcal{F}_5\right],\\
L_{l-}=&\int_{-1}^1\frac{dx}{2l}\left[P_{l-1}\mathcal{F}_6+P_l \mathcal{F}_5\right].
\end{split}
\label{eqn:multipole}
\end{equation}
   In the threshold region, the multipoles $\mathcal{M}_{l\pm}$ ($\mathcal{M}=E,M,L$) are proportional
to $|\vec{q}|^l$.
   To get rid of this purely kinematical dependence, one introduces reduced multipoles
$\overline{\mathcal{M}}_{l\pm}$ via
\begin{equation}
\overline{\mathcal{M}}_{l\pm}=\frac{\mathcal{M}_{l\pm}}{|\vec{q}|^l}.
\end{equation}

   Due to the assumed isospin symmetry, the process involves only three independent isospin structures
for the four physical channels \cite{Chew:1957tf}.
   Any amplitude $M$ for producing a pion with Cartesian isospin index
$a$ can be decomposed as
\begin{equation}
M(\pi^a)=\chi_f^\dagger\left(i\epsilon^{a3b}\tau^bM^{(-)}+\tau^aM^{(0)}+\delta^{a3}M^{(+)}\right)\chi_i,
\qquad a=1,2,3,
\label{eqn:isospin}
\end{equation}
where $\chi_i$ and $\chi_f$ denote the isospinors of the initial and final nucleons, respectively,
and $\tau^a$ are the Pauli matrices.
   The isospin amplitudes corresponding to $A_i$ of Eq.\ (\ref{eqn:dennery}) obey a crossing symmetry,
\begin{equation}
\begin{split}
A^{(0,+)}_i&\stackrel{s\leftrightarrow u}{\longrightarrow}\eta_i A^{(0,+)}_i,\\
A^{(-)}_i&\stackrel{s\leftrightarrow u}{\longrightarrow}-\eta_i A^{(-)}_i,
\end{split}
\label{eqn:crossing}
\end{equation}
where $\eta_i=1$ for $i=1,2,4$ and $\eta_i=-1$ for $i=3,5,6$.
    The physical reaction channels are related to the isospin channels via
\begin{equation}
\begin{split}
A_i(\gamma^{(*)}p\rightarrow n\pi^+)&=\sqrt{2}\left(A_i^{(-)}+A_i^{(0)}\right),\\
A_i(\gamma^{(*)}p\rightarrow p\pi^0)&=A_i^{(+)}+A_i^{(0)},\\
A_i(\gamma^{(*)}n\rightarrow p\pi^-)&=-\sqrt{2}\left(A_i^{(-)}-A_i^{(0)}\right),\\
A_i(\gamma^{(*)}n\rightarrow n\pi^0)&=A_i^{(+)}-A_i^{(0)}.
\end{split}
\label{eqn:isophysbez}
\end{equation}
   In the one-photon-exchange approximation, the differential cross section can be written as
\begin{equation}
\frac{d\sigma}{d\mathcal{E}_fd\Omega_fd\Omega_\pi^\textnormal{\,cm}}
=\Gamma\frac{d\sigma_v}{d\Omega_\pi^\textnormal{\,cm}},
\end{equation}
where the flux of the virtual photon is given by
\begin{equation}
\Gamma=\frac{\alpha}{2\pi^2}\frac{\mathcal{E}_f}{\mathcal{E}_i}\frac{k_\gamma}{Q^2}\frac{1}{1-\epsilon}.
\label{eq:Gamma}
\end{equation}
   In Eq.~(\ref{eq:Gamma}), $\mathcal{E}_i$ and $\mathcal{E}_f$ denote the energy of the initial and final electrons
in the laboratory frame, respectively, and $k_\gamma=(W^2-m_N^2)/(2m_N)$ is the so-called photon equivalent
energy in the laboratory frame.
   The parameter $\epsilon$ expresses the transverse polarization of the virtual
photon in the laboratory frame.
   In terms of laboratory electron variables it is given by
\begin{equation}
\epsilon=\left(1+2\frac{\vec{k}^2}{Q^2}\tan^2{\left(\frac{\Theta_e}{2}\right)}\right)^{-1},
\end{equation}
where $\Theta_e$ is the scattering angle of the electron.

   For an unpolarized target and without recoil polarization detection,
the virtual-photon differential cross section for pion production (subscript $v$)
can be further decomposed as \cite{Drechsel:1992pn}
\begin{equation}
\frac{d\sigma_v}{d\Omega_\pi}=\frac{d\sigma_T}{d\Omega_\pi}
+\epsilon\frac{d\sigma_L}{d\Omega_\pi}
+\sqrt{2\epsilon(1+\epsilon)}\frac{d\sigma_{LT}}{d\Omega_\pi}\cos\Phi_\pi
+\epsilon\frac{d\sigma_{TT}}{d\Omega_\pi}\cos{2\Phi_\pi}
+h\sqrt{2\epsilon(1-\epsilon)}\frac{d\sigma_{LT'}}{d\Omega_\pi}\sin{\Phi_\pi},
\label{eqn:wqparts}
\end{equation}
where it is understood that the variables of the individual virtual-photon cross sections
$d\sigma_T/d\Omega_\pi$ etc.~refer to the cm frame.
   For further details, especially concerning polarization observables,
we refer to Ref.\ \cite{Drechsel:1992pn}.
   The connection to the CGLN amplitudes can be found in Appendix \ref{ampcon}.
   Now we have all necessary formulas at hand to calculate pion electroproduction in an arbitrary
covariant and gauge-invariant framework.
   In the following section, we will introduce ChPT as an effective field theory which will allow us to
calculate pion production.
   The upper limit for the cm total energy $W$ is restricted by the fact that we consider pion and
nucleon degrees of freedom, only, and do not include the $\Delta(1232)$ resonance
\cite{Hemmert:1997ye,Hacker:2005fh,Pascalutsa:2005vq}.
   Furthermore, given the experience of the description of electromagnetic form factors, a
conservative/optimistic estimate for the upper limit of momentum transfers is $Q^2=0.1/0.2$ GeV$^2$.
   The inclusion of vector and axial-vector mesons leads to a much improved description
of the electromagnetic and axial form factors, respectively
\cite{Kubis:2000zd,Schindler:2005ke,Schindler:2006it,Bauer:2012pv}.

\section{Chiral perturbation theory}
   So far, an ab initio QCD calculation of electromagnetic pion production in the low-energy regime is not yet available.
   However, essential constraints of QCD, resulting from chiral symmetry, its spontaneous breakdown, and the explicit
breaking due to the quark masses, may be analyzed in terms of an effective field theory, namely,
chiral perturbation theory (see, e.g., Refs.~\cite{Ecker:1994gg,Bernard:1995dp,Scherer:2002tk,Scherer:2012zzd}
for an introduction).
   Starting point is a global $\mbox{SU}(2)_L\times\mbox{SU}(2)_R\times\mbox{U}(1)_V$ symmetry (chiral symmetry)
of QCD for massless $u$ und $d$ quarks, which is spontaneously broken down to $\mbox{SU}(2)_V\times
\mbox{U}(1)_V$ in the QCD ground state.
   In ChPT, the dynamics is expressed in terms of effective degrees of freedom (initially only pions,
subsequently also nucleons, etc.)
instead of the fundamental degrees of freedom of QCD (quarks and gluons).
   The purpose of ChPT is the construction of the most general theory describing the dynamics of the
Goldstone bosons driven by the underlying chiral symmetry of QCD.
   It was first developed for the mesonic sector of the lightest pseudoscalar mesons
\cite{Weinberg:1978kz,Gasser:1983yg}, as these are assumed to represent the Goldstone bosons associated
with the spontaneous symmetry breakdown in QCD.
   The pions can be described via the following unimodular unitary $(2\times 2)$ matrix,
\begin{equation}
\begin{split}
U(x)&=\textnormal{exp}\left(i\frac{\Phi(x)}{F}\right),\\
\Phi(x)&=\sum_{i=1}^3\tau_i \phi_i(x)=
\left(\begin{array}{cc}\pi^0(x) & \sqrt{2}\pi^+(x)\\ \sqrt{2}\pi^-(x)&-\pi^0(x)\end{array}\right),
\end{split}
\label{eqn:pionmatrix}
\end{equation}
where $F$ denotes the pion-decay constant in the chiral limit:
$F_\pi=F\left[1+O(\hat{m})\right]=92.2$ MeV with $\hat m=m_u=m_d$
being the isospin-symmetric limit of the light-quark masses.
   The most general effective Lagrangian is constructed in terms of $U$, covariant derivatives,
and external fields such that all desired symmetries are fulfilled.
   The external fields also allow one to systematically incorporate the consequences due to explicit
symmetry breaking in terms of the quark masses.
   This prescription, in principle, leads to a Lagrangian with an infinite number of terms, each
accompanied by a low-energy (coupling) constant (LEC).
   The complete mesonic Lagrangian can symbolically be written as
\begin{equation}
\mathcal{L}_\pi=\mathcal{L}_\pi^{(2)}+\mathcal{L}_\pi^{(4)}+\ldots,
\label{eq:Lpi}
\end{equation}
where the superscripts denote the chiral order (number of derivatives) of the Lagrangian.
   Physical observables are calculated perturbatively in terms of a quark-mass and momentum expansion.
   As one cannot make predictions by calculating an infinite number of diagrams,
Weinberg suggested a power counting scheme \cite{Weinberg:1978kz} which can be described as follows.
   Consider a given diagram calculated in the framework of Eq.~(\ref{eq:Lpi})
and re-scale the external momenta linearly, $p_i\mapsto t p_i$,
and the quark masses quadratically, $m_q\mapsto t^2 m_q$:
\begin{equation}
\mathcal{M}(tp_i,t^2m_q)=t^D\mathcal{M}(p_i,m_q).
\end{equation}
   The chiral dimension $D$ of the amplitude $\mathcal{M}$ estimates how important a diagram is
for the process at hand.
   The diagram is said to be of $O(q^D)$, where $q$ denotes a small momentum or a pion mass
and the property small refers to some scale of the order of 1 GeV.
   In $n$ dimensions, $D$ is given by
\begin{eqnarray}
\label{eq:D}
D&=&
2+(n-2)N_L+\sum_{k=1}^\infty 2(k-1)N_{2k}^\pi
\\
&\geq&2\,\,\textnormal{in four space-time dimensions,}\nonumber
\end{eqnarray}
where $N_L$ is the number of independent loops and
$N_{2k}^\pi$ the number of vertices from ${\cal L}_\pi^{2k}$.
   In particular, Eq.\ (\ref{eq:D}) establishes a relation
between the momentum and loop expansions, because at each chiral
order, the maximum number of loops is bounded from above.

   The lowest-order Lagrangian is given by \cite{Gasser:1983yg}
\begin{equation}
\mathcal{L}_\pi^{(2)}=\frac{F^2}{4}\textnormal{Tr}[D_\mu U(D^\mu U)^\dagger]
+\frac{F^2}{4}\textnormal{Tr}\left(\chi U^\dagger+U\chi^\dagger\right),
\end{equation}
where the covariant derivative $D_\mu U=\partial_\mu U-i r_\mu U+i U l_\mu$
contains the coupling to the external fields $r_\mu$ and $l_\mu$.
   The coupling to an external electromagnetic four-vector potential $\mathcal{A}_\mu$
is described by $r_\mu=l_\mu=-e\tau_3\mathcal{A}_\mu/2$.
   Furthermore, $\chi=2B(s+ip)$ includes the quark masses as
$\chi=2B\hat{m}=M^2$, where $M^2$ is the squared pion mass at leading order in the
quark-mass expansion
and $B$ is related to the scalar singlet
quark condensate $\left\langle \bar{q}q\right\rangle_0$
in the chiral limit \cite{Gasser:1983yg,Colangelo:2001sp}.
   The parameters $F$ and $B$ are the LECs of the leading-order Lagrangian.

   For the calculation of pion production at $O(q^4)$ we also need the next-to-leading-order
mesonic Lagrangian \cite{Gasser:1983yg,Gasser:1987rb},
\begin{eqnarray}
\mathcal{L}_\pi^{(4)}&=&\frac{l_3+l_4}{16}\left[\textnormal{Tr}(\chi U^\dagger+U\chi^\dagger)\right]^2
+\frac{l_4}{8}\textnormal{Tr}\left[D_\mu U(D^\mu U)^\dagger\right]
\textnormal{Tr}\left(\chi U^\dagger+U\chi^\dagger\right)\nonumber\\
&&+i\frac{l_6}{2}\textnormal{Tr}\left[f_{\mu\nu}^RD^\mu U(D^\nu U)^\dagger+f_{\mu\nu}^L(D^\mu U)^\dagger D^\nu U\right]
+\ldots,
\end{eqnarray}
where
\begin{equation}
\begin{split}
f^R_{\mu\nu}&=\partial_\mu r_\nu-\partial_\nu r_\mu-i [r_\mu,r_\nu],\\
f^L_{\mu\nu}&=\partial_\mu l_\nu-\partial_\nu l_\mu-i [l_\mu,l_\nu].
\end{split}
\label{eqn:mesonbausteine}
\end{equation}
   The $l_i$ are additional LECs and we have shown only the part of $\mathcal{L}_\pi^{(4)}$ relevant for
pion electroproduction.

   Besides the purely mesonic Lagrangian $\mathcal{L}_{\pi}$, we also need to discuss the part containing the
pion-nucleon interaction ($\mathcal{L}_{\pi N}$).
   For that purpose, let
\begin{equation}
\Psi=\left(\begin{array}{c}p\\n\end{array}\right)
\end{equation}
denote the nucleon field with two four-component Dirac fields for the proton and the neutron.
   Due to the spin-1/2 nature of the nucleon, the construction of $\mathcal{L}_{\pi N}$
also involves gamma matrices.
   Hence, additional building blocks appear in the construction of the Lagrangian.
   We refer the reader to Refs.\ \cite{Gasser:1987rb,Ecker:1995rk,Fettes:2000gb,Scherer:2012zzd} for further details.
   The lowest-order Lagrangian is given by \cite{Gasser:1987rb}
\begin{equation}
{\cal L}_{\pi N}^{(1)}=\bar{\Psi}\left(i\slashed{D}-m+\frac{\texttt{g}_A}{2}\gamma^\mu
\gamma_5 u_\mu \right)\Psi,\label{eq:LpiN}
\end{equation}
with
\begin{equation}
\begin{split}
D_\mu \Psi&=\left(\partial_\mu+\Gamma_\mu-iv_\mu^{(s)}\right)\Psi,\\
\Gamma_\mu&=\frac{1}{2}[u^\dagger(\partial_\mu-ir_\mu)u+u(\partial_\mu-il_\mu)u^\dagger],\\
u_\mu&=i[u^\dagger(\partial_\mu-ir_\mu)u-u(\partial_\mu-il_\mu)u^\dagger],\\
u&=\sqrt{U},
\end{split}
\label{eqn:nuktrafo}
\end{equation}
where $ v_{\mu}^{(s)}=-e \mathcal{A}_\mu/2$.
   In Eq.~(\ref{eq:LpiN}), the two LECs $m$ and $\texttt{g}_A$ denote the chiral limit
of the physical nucleon mass and the axial-vector coupling constant,
respectively.
   The expressions for the higher-order Lagrangians in the nucleon sector are lengthy
\cite{Gasser:1987rb,Ecker:1995rk,Fettes:2000gb}.
   Therefore, we focus only on the terms generating contact diagrams in pion photo- and
electroproduction.
  At ${O}(q^3)$, these terms read
\begin{equation}
\begin{split}
\mathcal{L}^{(3)}_{\pi N}&=\frac{d_8}{2m}
\left(i\bar{\Psi}\epsilon^{\mu\nu\alpha\beta}\textnormal{Tr}\left(\tilde{f}_{\mu\nu}^+u_\alpha\right)D_\beta\Psi
+\textnormal{H.c.}\right)\\
&\quad+\frac{d_9}{2m}\left(i\bar{\Psi}\epsilon^{\mu\nu\alpha\beta}
\textnormal{Tr}\left(f_{\mu\nu}^++2v_{\mu\nu}^{(s)}\right)u_\alpha D_\beta\Psi
+\textnormal{H.c.}\right)\\
&\quad-\frac{d_{20}}{8m^2}\left(i\bar{\Psi}\gamma^\mu\gamma_5\left[\tilde{f}_{\mu\nu}^+,u_\lambda
\right]D^{\lambda\nu}\Psi+\textnormal{H.c.}\right)\\
&\quad+i\frac{d_{21}}{2}\bar{\Psi}\gamma^\mu\gamma_5\left[\tilde{f}_{\mu\nu}^+,u^\nu\right]\Psi,
\end{split}
\label{eqn:lpn3}
\end{equation}
where H.c.\ refers to the Hermitian conjugate.
   The pion appears after expanding $u_\mu$, and
the photon is contained in the field-strength tensors
$f_{\mu\nu}^+$, $\tilde{f}_{\mu\nu}^+$, and $v_{\mu\nu}^{(s)}$.
   For further definitions, the reader is referred to Ref.\ \cite{Fettes:2000gb}.
   At order $O(q^4)$, the following additional interaction terms contribute to the contact graphs:
\begin{equation}
\begin{split}
\mathcal{L}^{(4)}_{\pi N}&=-\frac{e_{48}}{4m}\left(i\bar{\Psi}\textnormal{Tr}\left(f_{\lambda\mu}^+
+2v_{\lambda\mu}^{(s)}\right)h^\lambda_\nu\gamma_5\gamma^\mu D^\nu\Psi+\textnormal{H.c.}\right)\\
&\quad-\frac{e_{49}}{4m}\left(i\bar{\Psi}\textnormal{Tr}\left(f_{\lambda\mu}^+ +2v_{\lambda\mu}^{(s)}\right)
h^\lambda_\nu\gamma_5\gamma^\nu D^\mu\Psi+\textnormal{H.c.}\right)\\
&\quad+\frac{e_{50}}{24m^3}\left(i\bar{\Psi}\textnormal{Tr}\left(f_{\lambda\mu}^+ +2v_{\lambda\mu}^{(s)}\right)
h_{\nu\rho}\gamma_5\gamma^\lambda D^{\mu\nu\rho}\Psi+\textnormal{H.c.}\right)\\
&\quad-\frac{e_{51}}{4m}\left(i\bar{\Psi}u^\lambda
\left[D_\lambda,\textnormal{Tr}\left(f_{\mu\nu}^+ +2v_{\mu\nu}^{(s)}\right)\right]\gamma_5\gamma^\mu D^\nu\Psi
+\textnormal{H.c.}\right)\\
&\quad-\frac{e_{52}}{4m}\left(i\bar{\Psi}u_\mu \left[D^\lambda,\textnormal{Tr}
\left(f_{\lambda\nu}^+ +2v_{\lambda\nu}^{(s)}\right)\right]\gamma_5\gamma^\mu D^\nu\Psi+\textnormal{H.c.}\right)\\
&\quad-\frac{e_{53}}{4m}\left(i\bar{\Psi}u_\mu \left[D^\lambda,\textnormal{Tr}
\left(f_{\lambda\nu}^+ +2v_{\lambda\nu}^{(s)}\right)\right]\gamma_5\gamma^\nu D^\mu\Psi
+\textnormal{H.c.}\right)\\
&\quad-\frac{e_{67}}{4m}\left(i\bar{\Psi} \textnormal{Tr}\left(\tilde{f}_{\lambda\mu}^+h^\lambda_{\nu}\right)
\gamma_5\gamma^\mu D^\nu\Psi+\textnormal{H.c.}\right)\\
&\quad-\frac{e_{68}}{4m}\left(i\bar{\Psi} \textnormal{Tr}\left(\tilde{f}_{\lambda\mu}^+h^\lambda_{\nu}\right)
\gamma_5\gamma^\nu D^\mu\Psi+\textnormal{H.c.}\right)\\
&\quad+\frac{e_{69}}{24m^3}\left(i\bar{\Psi} \textnormal{Tr}
\left(\tilde{f}_{\lambda\mu}^+h_{\nu\rho}\right)\gamma_5\gamma^\lambda D^{\mu\nu\rho}\Psi
+\textnormal{H.c.}\right)\\
&\quad-\frac{e_{70}}{4m^2}\left(i\bar{\Psi}\left[\tilde{f}_{\lambda\mu}^+,h_{\nu\rho}\right]
{\epsilon^{\lambda\mu\nu}}_\tau D^{\rho\tau}\Psi+\textnormal{H.c.}\right)\\
&\quad-\frac{e_{71}}{4m}\left(i\bar{\Psi} \textnormal{Tr}\left(u^\lambda
\left[D_\lambda,\tilde{f}_{\mu\nu}^+\right]\right)\gamma_5\gamma^\mu D^{\nu}\Psi
+\textnormal{H.c.}\right)\\
&\quad-\frac{e_{72}}{4m}\left(i\bar{\Psi} \textnormal{Tr}
\left(u_\mu\left[D^\lambda,\tilde{f}_{\lambda\nu}^+\right]\right)\gamma_5\gamma^\mu D^{\nu}\Psi
+\textnormal{H.c.}\right)\\
&\quad-\frac{e_{73}}{4m}\left(i\bar{\Psi} \textnormal{Tr}
\left(u_\mu\left[D^\lambda,\tilde{f}_{\lambda\nu}^+\right]\right)\gamma_5\gamma^\nu D^{\mu}\Psi
+\textnormal{H.c.}\right)\\
&\quad-\frac{e_{112}}{4m}\left(\bar{\Psi}\textnormal{Tr}
\left(f_{\mu\nu}^++2v_{\mu\nu}^{(s)}\right)\tilde{\chi}_-\gamma_5\gamma^\mu D^\nu\Psi
+\textnormal{H.c.}\right)\\
&\quad-\frac{e_{113}}{4m}\left(\bar{\Psi}\textnormal{Tr}
\left(\tilde{f}_{\mu\nu}^+\tilde{\chi}_-\right)\gamma_5\gamma^\mu D^\nu\Psi
+\textnormal{H.c.}\right).
\end{split}
\label{eqn:lpn4}
\end{equation}
   In the present calculation, the Lagrangians of Eqs.\ (\ref{eqn:lpn3}) and (\ref{eqn:lpn4}) will be used at tree
level only.
   In a calculation at $O(q^4)$, we can replace $m$ by $m_N$, because the difference is of $O(q^2)$
and will first show up at $O(q^5)$.

   In the single-nucleon sector, the power-counting formula of Eq.~(\ref{eq:D}) is modified according to
\cite{Ecker:1994gg}
\begin{align}
\label{eq:DN}
D&=1+(n-2)N_L+\sum_{k=1}^\infty 2(k-1)N_{2k}^\pi+\sum_{k=1}^\infty (k-1)N_{k}^N\\
&\geq 1\,\,\textnormal{in four space-time dimensions,}\nonumber
\end{align}
where $N_k^N$ is the number of vertices derived from ${\cal L}_{\pi N}^{(k)}$.
   When the methods of mesonic ChPT were applied to the one-nucleon sector for the first
time, it was noted that loop diagrams contributed to lower orders than predicted by the
power counting \cite{Gasser:1987rb}.
   In other words, the correspondence between the chiral expansion and the loop expansion was
seemingly lost.
   It was also noted that the violation of the power counting was due to applying dimensional regularization in
combination with the modified minimal subtraction scheme of ChPT to loop diagrams.
   The infrared renormalization of Ref.\ \cite{Becher:1999he} and the
extended on-mass-shell (EOMS) scheme of Refs.\ \cite{Gegelia:1999gf,Fuchs:2003qc}
addressed this problem in a manifestly Lorentz-invariant framework.
   It was shown that the power-counting-violating terms can be absorbed through a
redefinition of the LECs such that the renormalized diagrams satisfy the power counting
of Eq.~(\ref{eq:DN}).
   Here, we exploit the results of the EOMS scheme in a somewhat modified manner.
   To be specific, the $O(q^3)$ LECs $d_8$, $d_9$, $d_{20}$, and $d_{21}$ have been
adjusted numerically without explicitly separating the power-counting-violating part (the details have already
been described in Appendix B of Ref.\ \cite{Hilt:2013uf} and will not be repeated here).

\section{Calculation of the matrix element}
   We have calculated the matrix element of Eq.~(\ref{eqn:ball}) up to and including $O(q^4)$
in the framework of manifestly Lorentz-invariant baryon chiral perturbation theory.
   The topologies for the one-loop diagrams were already listed in Ref.\ \cite{Bernard:1992nc}.
   In an effective field theory, every diagram has multiple contributions, where only the structure of
the vertices changes.
   In the present case, the same vertex can have different chiral orders.
   Hence, up to the accuracy we are working, there exist 85 loop and 20 tree diagrams.
   Calculating these diagrams is fairly straightforward but cumbersome because of the size of the expressions
involved.
   We therefore used the computer algebra system {\small  MATHEMATICA} together with the
{\small FEYNCALC package} \cite{Mertig:1990an} to calculate the diagrams.
   Nevertheless, the final result needs to be checked.
   We have explicitly verified that current conservation, Eqs.\ (\ref{eqn:stromerhalt2}), and crossing symmetry,
Eqs.\ (\ref{eqn:crossing}), are fulfilled analytically for our results.
   To evaluate loop integrals, we made use of the LoopTools package \cite{Hahn:2000kx}.

   In Table \ref{tab:lecs}, we display LECs of $\mathcal{L}_{\pi}$ and $\mathcal{L}_{\pi N}$
which have been extracted from processes other than pion photo- and electroproduction,
such as form factors of the nucleon and the pion.
   On the other hand, all LECs entering only the contact diagrams resulting from the Lagrangians
of Eqs.\ (\ref{eqn:lpn3}) and (\ref{eqn:lpn4}) are determined in fits to pion production data.
   The details of this procedure are the subject of the next section.

\begin{table}[t]
            \caption{LECs determined from other processes.}
    \label{tab:lecs}
    \centering
        \begin{tabular}{ll}
        \hline
        \hline
    LEC       &     Source\\
    \hline
    $l_3$     & $M_\pi=134.977$ MeV \cite{Beringer:1900zz}\\
    $l_4$, $l_6$ & pion form factor \cite{Bijnens:1998fm}\\
    $c_1$     &      proton mass $m_p=938.272$ MeV   \cite{Beringer:1900zz}        \\
    $c_2$, $c_3$, $c_4$ & pion-nucleon scattering \cite{Becher:2001hv}\\
    $c_6$, $c_7$ &  magnetic moment of proton ($\mu_p=2.793$) and neutron ($\mu_n=-1.913$) \cite{Beringer:1900zz}    \\
    $d_6$, $d_7$, $e_{54}$, $e_{74}$  & world data for nucleon electromagnetic form factors
    ($Q^2<0.3$ $\textnormal{GeV}^2$) \cite{merkel}     \\
   $d_{16}$  &  axial-vector coupling constant $g_A=1.2695$  \cite{Beringer:1900zz}      \\
    $d_{18}$  &  pion-nucleon coupling constant\footnote{\,
    In Ref.\ \cite{Baru:2010xn}, the value of the charged-pion-nucleon coupling constant
    was extracted to be $g_c^2/(4\pi)=13.69\pm 0.20$.} $g_{\pi N}=13.21$ \cite{Schroder:2001rc}\\
    $d_{22}$  &  axial radius of the nucleon $\langle r_A^2\rangle =12/M_A^2$, $M_A=1.026$ GeV \cite{Liesenfeld:1999mv} \\
        \hline
        \hline
        \end{tabular}
\end{table}

\section{Determination of low-energy constants}
   At $O(q^3)$, four independent LECs exist [see Eq.~(\ref{eqn:lpn3})] which are specifically related
to pion photoproduction.
   Two of them, $d_{20}$ and $d_{21}$, enter the isospin ($-$) channel and are, therefore, only relevant
for the production of charged pions.
   Moreover, they contribute differently to the invariant amplitudes $A_i$ of Eq.~(\ref{eqn:dennery}).
   The remaining two constants $d_8$ and $d_9$ enter the isospin
(+) and (0) channels, respectively, though both in combination with the same Dirac structure.
   Finally, at $O(q^3)$ the description of pion electroproduction is a prediction,
because no new parameter (LEC) beyond photoproduction is available at that order.

   At $O(q^4)$,  15 additional LECs appear [see Eq.~(\ref{eqn:lpn4})].
   In the case of pion photoproduction, the five constants $e_{48}$ --
$e_{51}$ and $e_{112}$ contribute to the isospin (0) channel, the five constants $e_{67}$ --
$e_{69}$, $e_{71}$, and $e_{113}$ to the isospin (+) channel, and the constant $e_{70}$ to the
isospin ($-$) channel.
   For electroproduction, the (0) and (+) channels each have two more independent LECs
$e_{52}$, $e_{53}$ and $e_{72}$, $e_{73}$, respectively.
   We note that the isospin $(-)$ channel, even at $O(q^4)$, does not contain any
free LEC specifically related to electroproduction.

   Now, how can one determine these LECs?
   Since the LECs parametrize the dynamics of the underlying fundamental theory, namely QCD,
they can, in principle, be obtained from lattice QCD.
   At present, however, the LECs of pion production are not available.
   Therefore, we focus on a determination in terms of fitting to experimental data,
where the accuracy depends on the amount and quality of the available data in the
various reaction channels.
   In this context, one has to determine the energy range in which ChPT can be applied.
   Initially, ChPT was constructed for the low-energy regime and, therefore,
it is particularly suited for the threshold region of pion production.
   Nevertheless, the situation turns out to be quite different for neutral pion production
in comparison with charged pion production.
   Predictions for the latter are rather precise even at lowest order which
is due to the Kroll-Ruderman theorem \cite{Kroll:1953vq}.
   The neutral channels are much more involved.
   There, the breaking of isospin symmetry plays a crucial role.
   This can be seen experimentally in the cusp in the $E_{0+}$ multipole
\cite{Faldt:1979fs,Bernstein:1998ip}.
   Theoretically it stems from the fact that, within a loop, in principle, either a proton and
the appropriate pion or a neutron and the appropriate pion can propagate.
   Both cases contribute to the amplitude but this effect is of higher order in an
$O(q^4)$ calculation.
   In  Ref. \cite{Bernard:1993bq} the effect was phenomenologically included by using
the mass of the $\pi^\pm$ within the loops.
   Here, we also exploit this idea.
   We consistently use $M_{\pi^0}$ and $m_p$ for mass parameters in the amplitudes and
$M_{\pi^\pm}$ and $m_n$ for the mass parameters in loop integrals.

   The fits we performed are of a nonlinear type in the parameters, because the observables are typically
proportional to the squared invariant amplitude.
   We therefore did several thousand fits with different starting values to make sure
that we found not only a local but the global minimum of the reduced $\chi^2_\textnormal{red}$.
   In order to estimate the errors of our parameters, we used the so-called bootstrap method \cite{efron}.
   The idea is as follows.
   Assuming a data set $\mathbf{Y}=y_1,\ldots,y_n$ of length $n$, one can create $m$
bootstrap samples $\mathbf{Y}_1,\ldots,\mathbf{Y}_m$ of length $n$, where $m$ should be a
sufficiently large number.
   The data points are randomly chosen to create the new data sets, where some data points now
appear several times and others are neglected.
   Every sample is fitted in the same way as the original data.
   In the end one has $m$ values for the parameters.
   According to the bootstrap method, the standard deviation of the $m$ values for each parameter
is an estimate for its error.
   Below, we discuss details for all reaction channels that were analyzed.

\subsection{$\gamma+p\rightarrow p+\pi^0$\label{pi0photo}}
   This reaction channel, including the electroproduction case to be
discussed in the next subsection, is particularly interesting, because the
leading-order term of the threshold production amplitude is predicted to be
zero due to the Kroll-Ruderman theorem \cite{Kroll:1953vq}.
   The latest experiment at the Mainz Microtron \cite{Hornidge:2012ca} was designed to analyze
the $P$ waves in the threshold region with very high precision.
   Therefore, not only differential cross sections but also the polarized
photon asymmetry $\Sigma$ (see Appendix \ref{diffcs}) has been measured.
   In Ref.\ \cite{Hilt:2013uf} we already discussed our results for this
channel in great detail.
   Here, we only summarize our findings.

   In the past, an analysis of $\pi^0$ photoproduction near threshold only involved
$S$ and $P$ waves.
   As was shown in Refs.\ \cite{FernandezRamirez:2009su,FernandezRamirez:2009jb},
$D$ waves are also very important, as they strongly influence the extraction of
other multipoles through interference with large $P$ waves.
   Hence, we used $S$, $P$, and $D$ waves to calculate the observables.
   This means we had to determine six independent LECs.
   In HBChPT, one can rearrange these constants such that two appear in $E_{0+}$
and one in every $P$ wave.
   Previously, the $D$ waves have not been analyzed and so the sixth
independent LEC has never been taken into account.
   In RChPT, the situation is more involved.
   One cannot rearrange the LECs in such a way as in HBChPT as there always appear
new mixings of the constants in the multipoles at higher order in a $1/m_N$ expansion
(see Appendix A of Ref.\ \cite{Hilt:2013uf}).
   Expanding the contact terms of the relativistic result in $1/m_N$,
one obtains for the leading-order term the same result as in HBChPT.

   Nevertheless, our fits of the experimental data showed that we could
not determine the sixth LEC denoted by $\tilde{e}_{49}=e_{49}+e_{68}$.
   The problem is that, even though
$D$ waves are important, they could not be separated in the current
experiment \cite{Hornidge:2012ca}.
   Hence, we neglected this LEC.
   We also neglected the first two energy bins below the $\pi^+$ threshold
($E_\gamma^\textnormal{lab}=147$ MeV and $E_\gamma^\textnormal{lab}=149.3$ MeV)
in our fits, as the experiment of Ref.\ \cite{Hornidge:2012ca}
was not particularly designed for energies below the $\pi^+$ threshold.
   This region is covered more precisely by other experiments
\cite{Fuchs:1996ja,Bergstrom:1996fq,Schmidt:2001vg}.
   Furthermore, below this threshold the $E_{0+}$ multipole, which dominates there,
is strongly constrained by unitarity.
   Since the data of Ref.~\cite{Hornidge:2012ca} were taken over a much wider energy range
than ChPT can be applied to, we had to determine the best energy range for a fit.
   Former results of HBChPT already indicated an upper limit of $E_\gamma^\textnormal{lab}<170$
MeV.
   We used the reduced $\chi^2_\textnormal{red}$ as an estimator for the energy
region to fit.
\begin{figure}[tbp]
    \centering
        \includegraphics[width=0.5\textwidth]{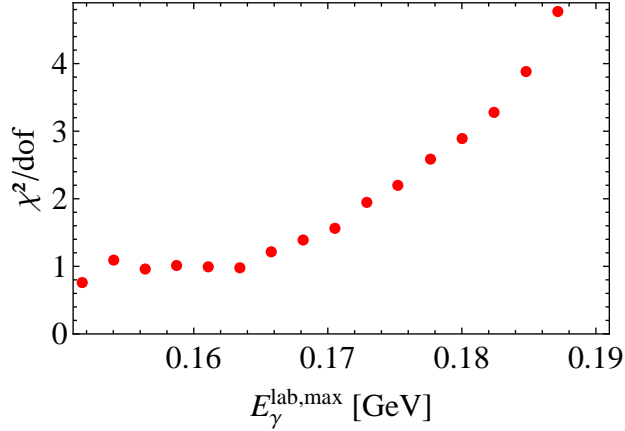}
    \caption{$\chi^2_{\textnormal{red}}$ as a function of the fitted energy range from
    $\pi^+$ threshold up to $E_\gamma^\textnormal{lab,max}$.}
    \label{fig:chi2plot}
\end{figure}
   In Fig.\ \ref{fig:chi2plot}, we show how the $\chi^2_\textnormal{red}$ changes if one includes all
data points up to a maximal energy $E_\gamma^\textnormal{lab,max}$.
   It stays around 1 up to bin 8 (bins 1 and 2 corresponding to $E_\gamma^\textnormal{lab}=147$
MeV and 149.3 MeV, respectively, not shown) and then starts to rise.
   Furthermore, we took account of the change of the LECs, when including higher energy bins.
   We decided to take all data up to the first rising bin, namely $E_\gamma^\textnormal{lab}=165.8$ MeV with
$\chi^2_\textnormal{red}=1.22$.
   Our results for the LECs, including an error estimate, are shown in Table \ref{tab:lecspi0p}.
\begin{table}[ht]
\caption{LECs of the contact diagrams for $\gamma+p\rightarrow p+\pi^0$ as obtained from
a fit to the data of Ref.~\cite{Hornidge:2012ca}. The $d_i$ are given in units of $\textnormal{GeV}^{-2}$ and
the $e_i$ in units of $\textnormal{GeV}^{-3}$. The errors stem from a bootstrap estimate (see text for details).}
    \label{tab:lecspi0p}
    \centering
        \begin{tabular}{cc}
        \hline
        \hline
    LEC       &     Value\\
    \hline
$\tilde{d}_9:=d_8+d_9$          &  $-2.31\pm0.02$  \\
$\tilde{e}_{48}:=e_{48}+e_{67}$      &   $-3.0\pm0.2$     \\
$\tilde{e}_{49}:=e_{49}+e_{68}$      &   0    \\
$\tilde{e}_{50}:=e_{50}+e_{69}$      &   $-1.2\pm2.1$   \\
$\tilde{e}_{51}:=e_{51}+e_{71}$      &   $2.3\pm1.1$   \\
$\tilde{e}_{112}:=e_{112}+e_{113}$   &   $-4.4\pm2.1$   \\
        \hline
        \hline
        \end{tabular}
\end{table}

\begin{figure}[tbp]
    \centering
        \includegraphics[width=\textwidth]{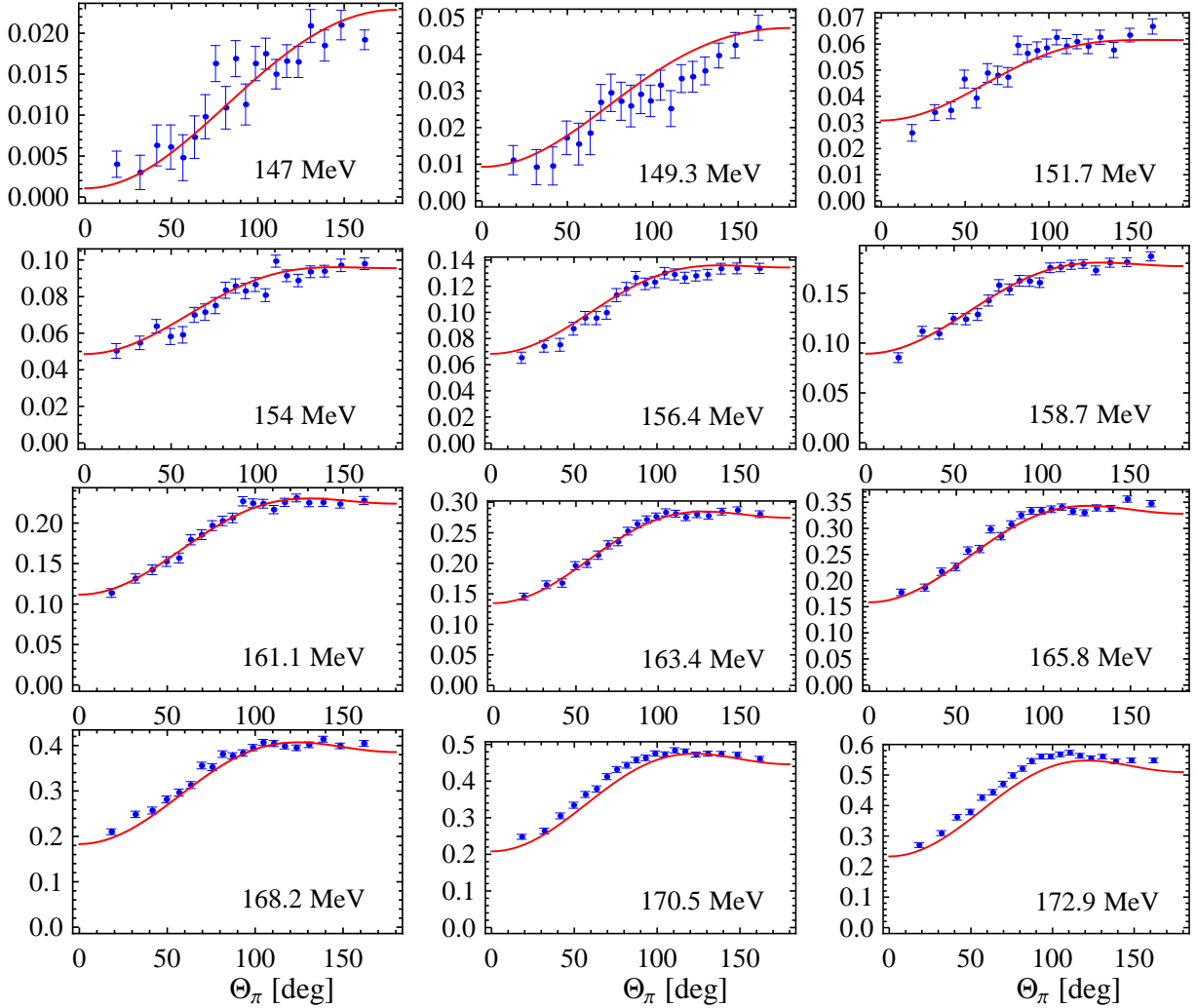}
    \caption{Angular distribution for differential cross sections in $\mu$b/sr for $\gamma+p\rightarrow p+\pi^0$.
    The curves show the results in RChPT at $O(q^4)$.
    The data are from Ref.\ \cite{Hornidge:2012ca}. The energy $E_\gamma^\textnormal{lab}$ is given in the panels.}
    \label{fig:wqgesamtp0photobernstein}
\end{figure}
\begin{figure}[tbp]
    \centering
        \includegraphics[width=\textwidth]{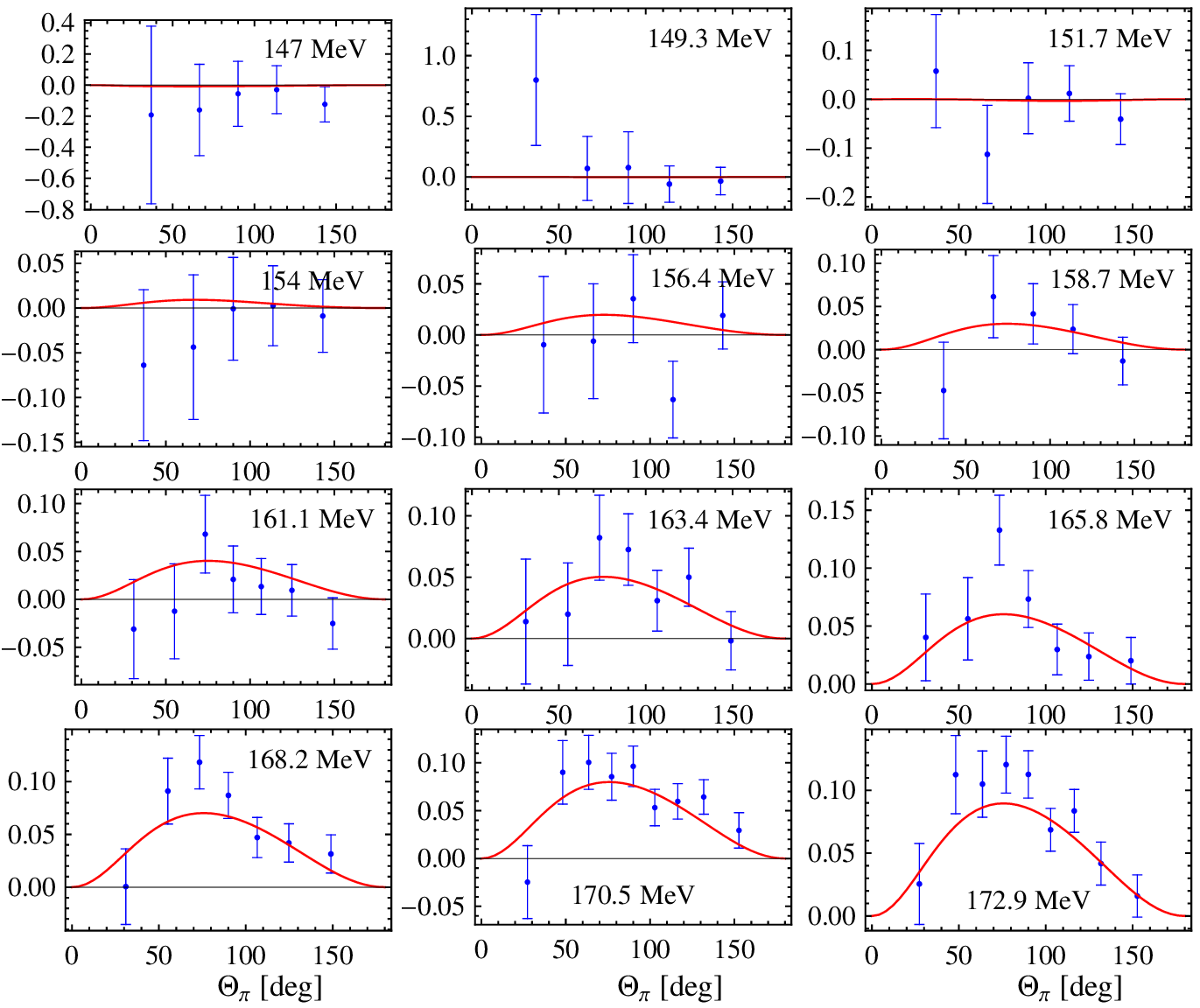}
    \caption{Angular distribution for the polarized photon asymmetries $\Sigma$ for $\gamma+p\rightarrow p+\pi^0$.
    The curves show the results in RChPT at $O(q^4)$. The data are from Ref.\ \cite{Hornidge:2012ca}.
The energy $E_\gamma^\textnormal{lab}$ is given in the panels.}
    \label{fig:sigmagesamtp0photobernstein}
\end{figure}
\begin{figure}[htbp]
    \centering
        \includegraphics[width=\textwidth]{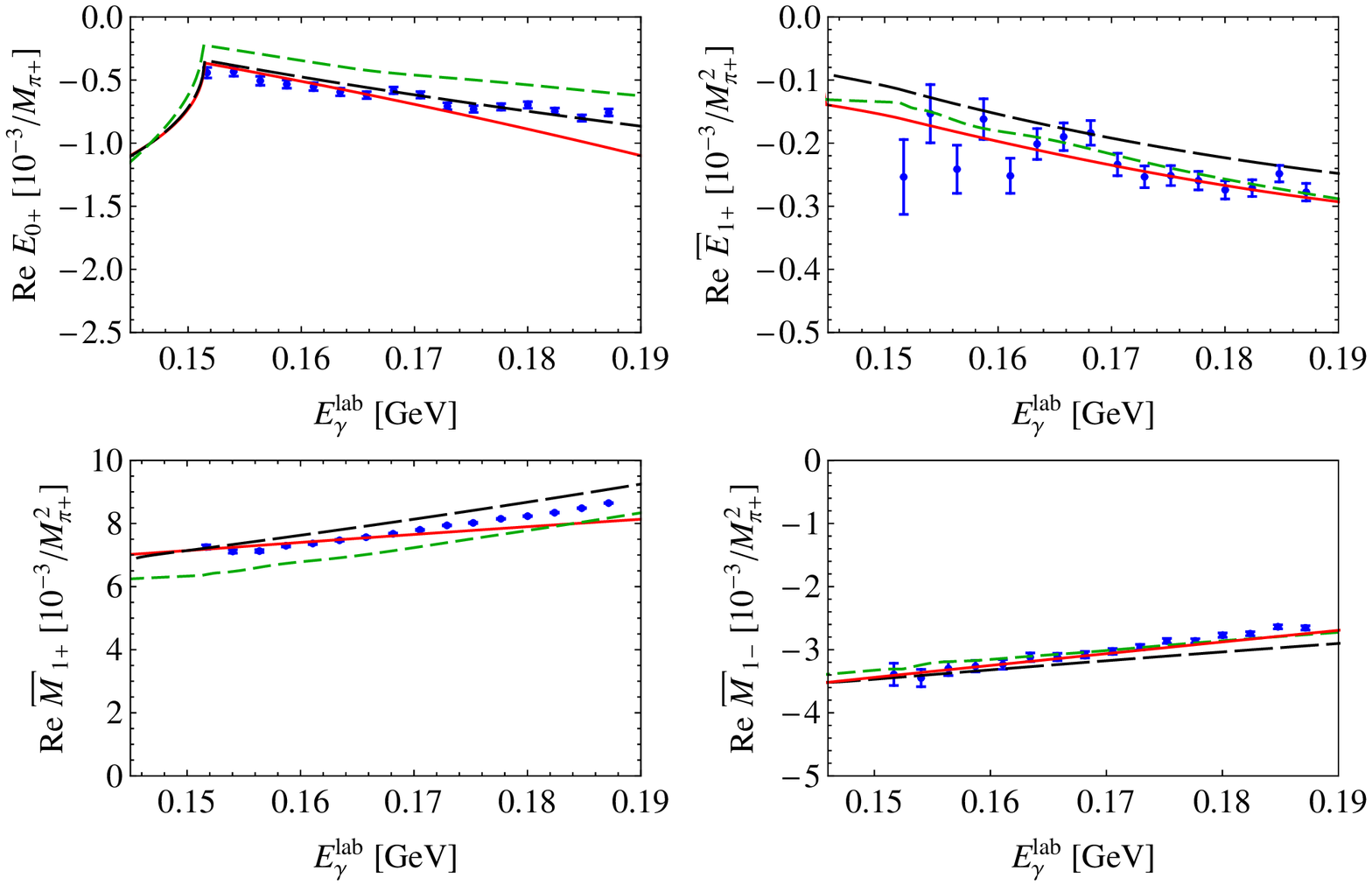}
    \caption{(Color online) $S$- and reduced $P$-wave multipoles for $\gamma+p\rightarrow p+\pi^0$.
    The solid (red) curves show our RChPT calculations at $O(q^4)$.
    The short-dashed (green) and long-dashed (black) curves are
    the predictions of the DMT model \cite{Kamalov:2000en,Kamalov:2001qg} and the GL model
    \cite{Gasparyan:2010xz}, respectively.
    The data are from Ref.\ \cite{Hornidge:2012ca}.}
    \label{fig:multipolesexportphyspi0p}
\end{figure}

   Some exemplary results for the differential cross section and the polarized photon asymmetry
are shown in Figs.\ \ref{fig:wqgesamtp0photobernstein} and \ref{fig:sigmagesamtp0photobernstein}.
   In the fitted energy range we get a nice agreement with the data.
   At higher energies, the calculation starts to deviate from the experiment, because the important
$M_{1+}$ multipole, which is dominated by the $\Delta$ resonance,
is underestimated.
   The real parts of the $S$ and $P$ waves are shown in Fig.\ \ref{fig:multipolesexportphyspi0p} together
with single-energy fits of Ref.\ \cite{Hornidge:2012ca}.
   For comparison, we also show the predictions
of the Dubna-Mainz-Taipei (DMT) model \cite{Kamalov:2000en,Kamalov:2001qg}
and the covariant, unitary, chiral approach of Gasparyan and Lutz (GL) \cite{Gasparyan:2010xz}.
   The multipole $E_{0+}$ agrees nicely with the data in the fitted energy range.
   The $P$ waves $E_{1+}$ and $M_{1-}$ agree for even higher energies with the
single energy fits.
   The largest deviation can be seen in $M_{1+}$.
   This multipole is related to the $\Delta$ resonance and the strong
rising of the data above 170 MeV can be traced back to the influence
of this resonance.
   As we did not include the $\Delta$ explicitly, this
calculation is not able to fully describe its impact on the multipole.

\subsection{$\gamma^{*}+p\rightarrow p+\pi^0$}

   After having fixed the LECs of $\pi^0$ photoproduction, there remain only two independent structures for
electroproduction.
   We use the latest data of Ref.\ \cite{Merkel:2011cf} to determine the corresponding LECs.
   In Ref.\ \cite{Merkel:2011cf}, the differential cross sections $\sigma_0=\sigma_T+\epsilon\sigma_L$
and $\sigma_{LT}$ were precisely measured in the threshold region for different values of $Q^2$.
   We use the same fitting procedure as in photoproduction and also apply the bootstrap method to
estimate the errors of the LECs (see Table \ref{tab:lecvaluespi0electro}).
\begin{table}[htbp]
\caption{\label{tab:lecvaluespi0electro}
LECs of the contact diagrams for $\gamma^\ast+p\rightarrow p+\pi^0$ as obtained from
a fit to the data of Ref.~\cite{Merkel:2011cf}. The $e_i$ are given in units
of $\textnormal{GeV}^{-3}$. The errors stem from a bootstrap estimate (see text for details).}
\centering
\begin{tabular}{cc}
\hline
\hline
LEC & Value\\
\hline
$\tilde{e}_{52}=e_{52}+e_{72}$ & $6.4\pm0.7$\\
$\tilde{e}_{53}=e_{53}+e_{73}$ & $-0.5\pm0.2$\\
\hline
\hline
\end{tabular}
\end{table}
   We obtain $\chi^2_{\textnormal{red}}=1.97$ as the global minimum.

\begin{figure}[htbp]
    \centering
        \includegraphics[width=\textwidth]{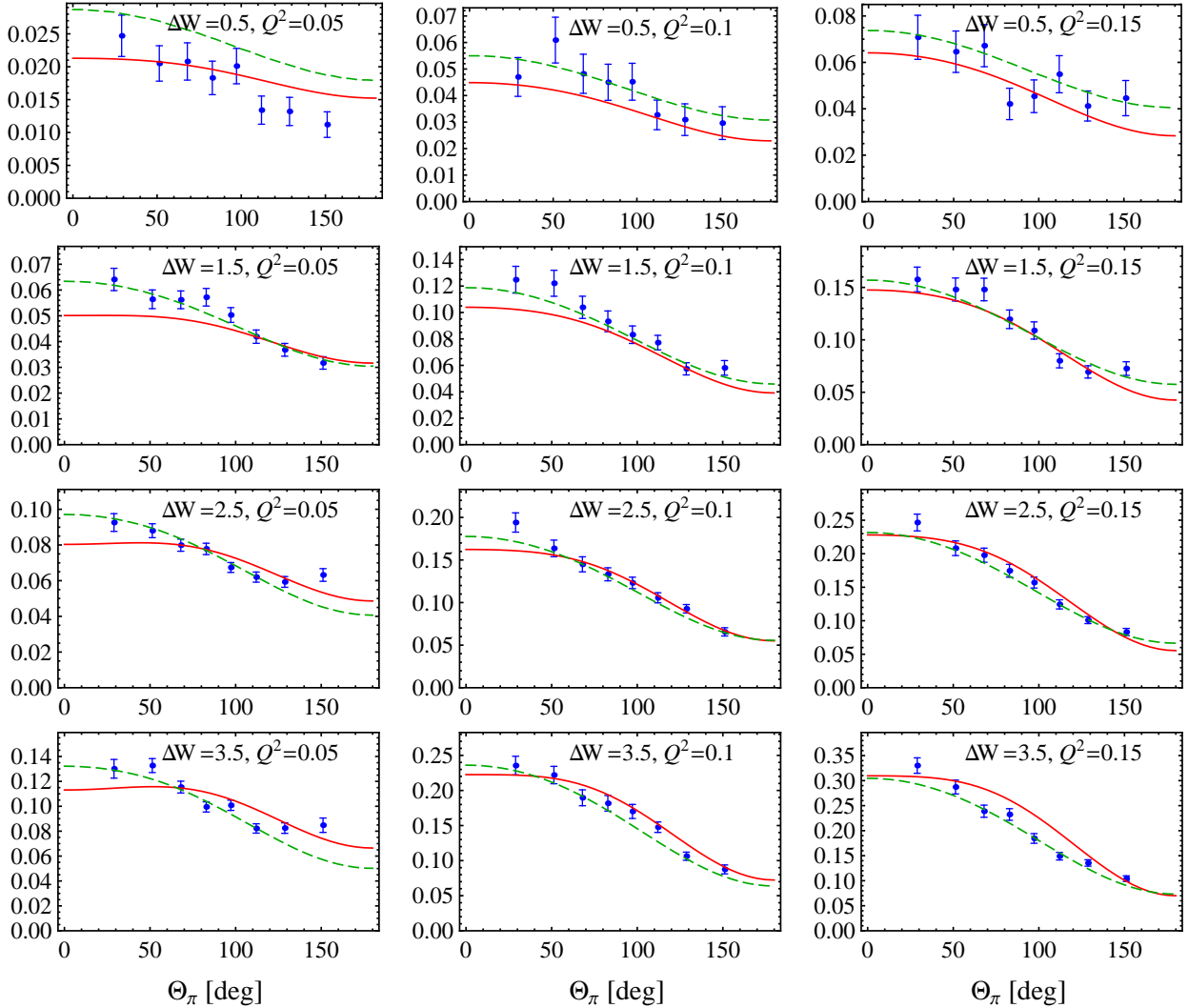}
    \caption{(Color online) Angular distribution for differential cross sections $\sigma_0$ in $\mu$b/sr for
    $\gamma^*+p\rightarrow p+\pi^0$.
    The values for the virtual-photon polarization $\epsilon$ are 0.932, 0.882, and 0.829 for increasing $Q^2$.
    The solid (red) curves show our RChPT calculations at $O(q^4)$.
    The short-dashed (green) curves are the predictions of the DMT model \cite{Kamalov:2000en,Kamalov:2001qg}.
    The data are from Ref.\ \cite{Merkel:2011cf}. The cm energy above threshold $\Delta W$
    and the photon virtuality $Q^2$ are given in the panels in units of MeV and GeV$^2$, respectively.}
    \label{fig:wqpi0electro}
\end{figure}

\begin{figure}[htbp]
    \centering
        \includegraphics[width=\textwidth]{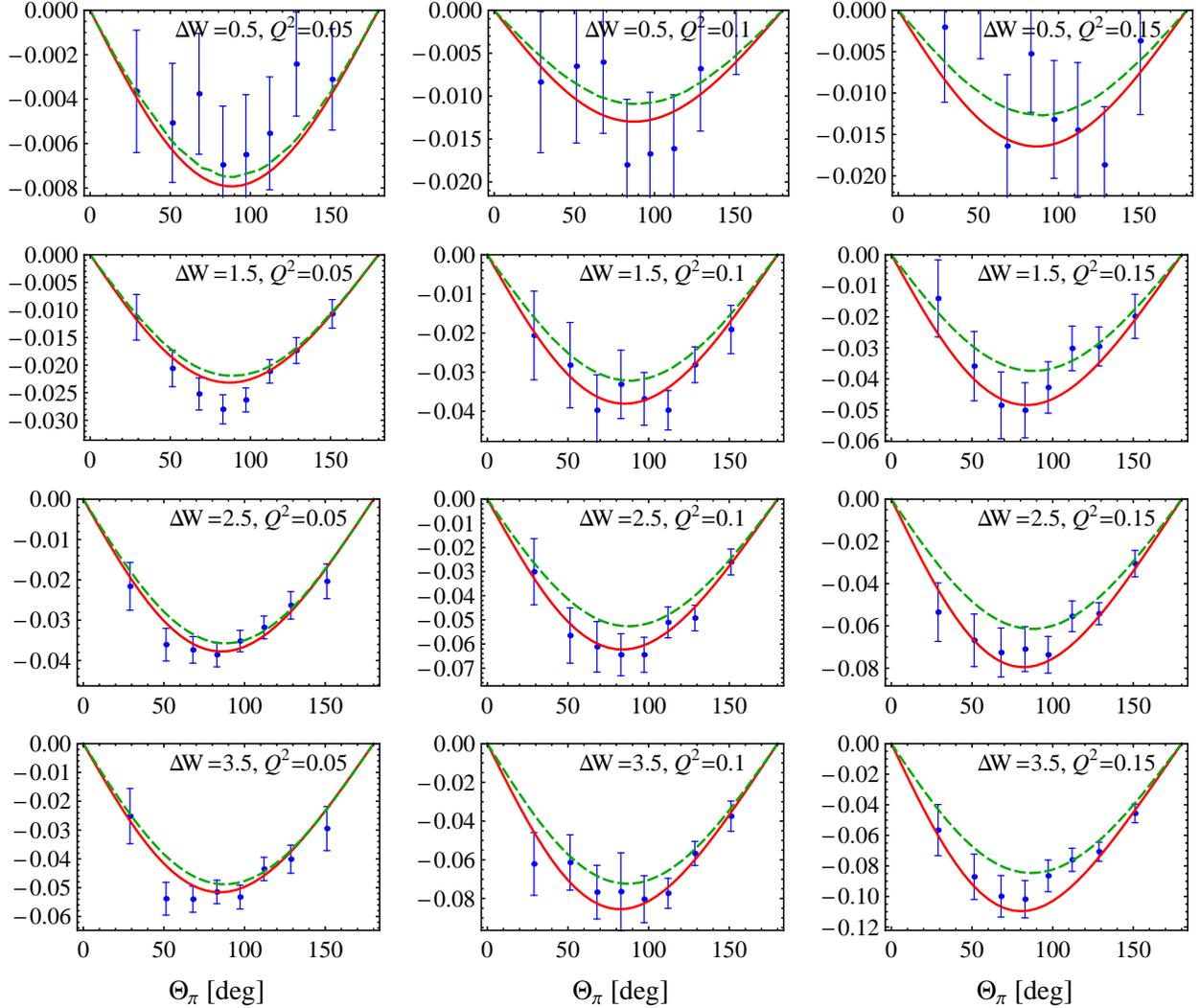}
    \caption{(Color online) Angular distribution for the differential cross sections $\sigma_{LT}$ in $\mu$b/sr
    for $\gamma^*+p\rightarrow p+\pi^0$.
    The curves and the polarization values $\epsilon$ are as in Fig.~\ref{fig:wqpi0electro}.
    The data are from Ref.\ \cite{Merkel:2011cf}.
    The cm energy above threshold $\Delta W$
    and the photon virtuality $Q^2$ are given in the panels in units of MeV and GeV$^2$, respectively.}
    \label{fig:sltpi0electro}
\end{figure}

\begin{figure}[htbp]
    \centering
        \includegraphics[width=0.80\textwidth]{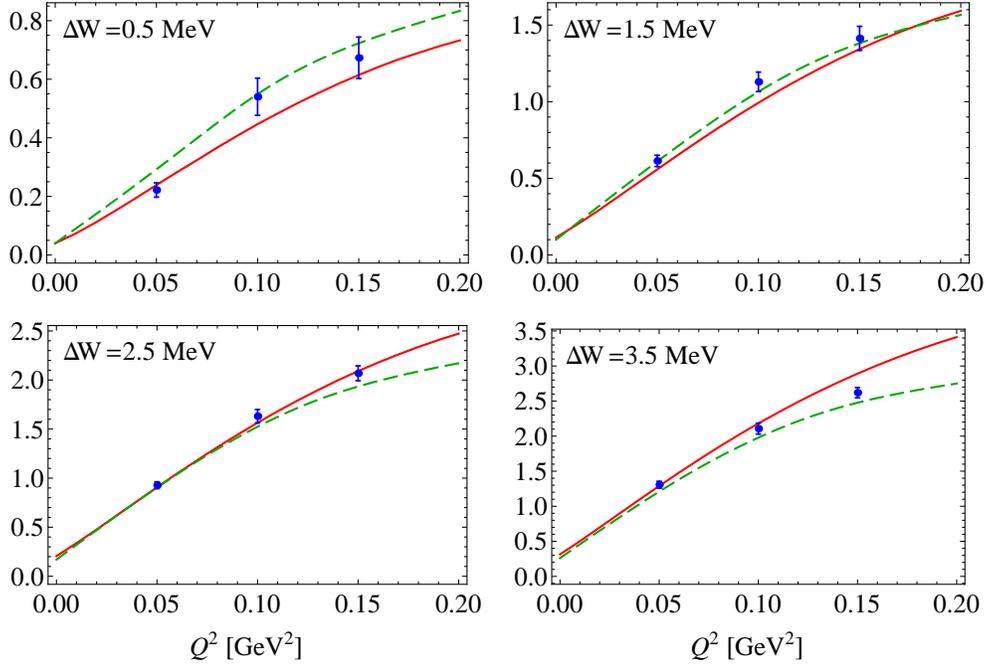}
    \caption{(Color online) Total cross sections in $\mu$b as a function of $Q^2$ for different cm energies
above threshold $\Delta W$ in MeV.
    The curves and the polarization values $\epsilon$ are as in Fig.~\ref{fig:wqpi0electro}.
   The data are from Refs.\ \cite{Merkel:2009zz,Merkel:2011cf}.}
    \label{fig:totcspi0electro}
\end{figure}

\begin{figure}[htbp]
    \centering
        \includegraphics[width=0.90\textwidth]{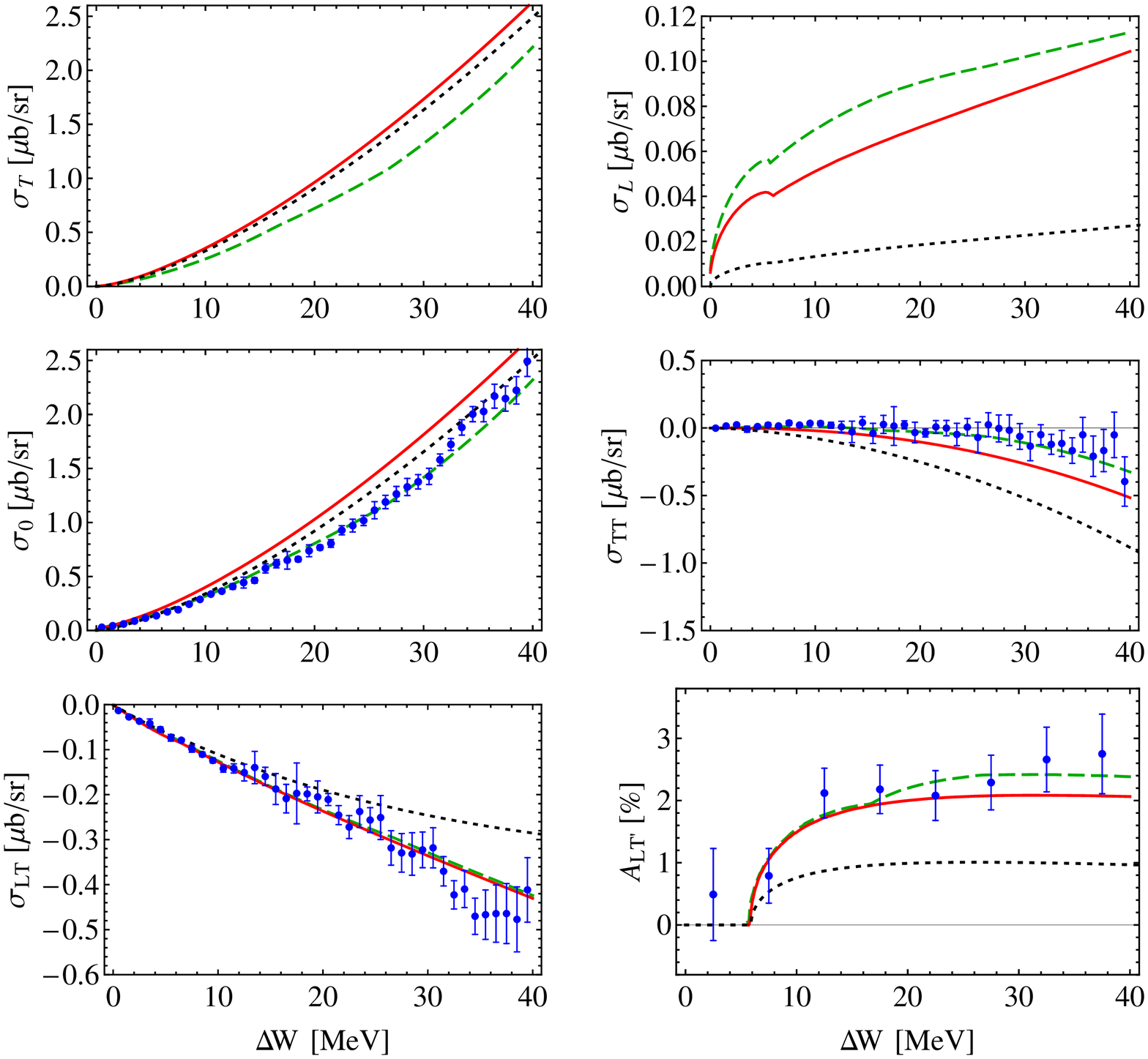}
    \caption{(Color online) Coincidence cross sections $\sigma_0$, $\sigma_{TT}$, and $\sigma_{LT}$
    in $\mu$b/sr and beam asymmetry
    $A_{LT'}$ in $\%$ at constant $Q^2=0.05$~GeV$^2$,
    $\Theta_\pi=90^\circ$, $\Phi_\pi=90^\circ$, and $\epsilon=0.93$  as a function of
    $\Delta W$ above threshold.
    The solid (red) lines show our RChPT calculations at $O(q^4)$ and the dotted (black) lines are the
    heavy-baryon ChPT calculations of Ref.~\cite{Bernard:1996ti}.
    The dashed (green) curves are obtained from the DMT
    model \cite{Kamalov:2000en,Kamalov:2001qg}.
    The data are from Ref.\ \cite{Weis:2007kf}.}
    \label{fig:weispi0electro}
\end{figure}

   The results for the differential cross sections are shown in Figs.\ \ref{fig:wqpi0electro}
and \ref{fig:sltpi0electro}.
   The calculation agrees nicely with the data.
   Furthermore, in Fig.\ \ref{fig:totcspi0electro} we show the total cross section for these
energies together with the experimental data \cite{Merkel:2009zz,Merkel:2011cf}.
   Finally, in Fig.~\ref{fig:weispi0electro} we compare our results for the coincidence
cross sections $\sigma_0$, $\sigma_{TT}$, $\sigma_{LT}$ and the beam asymmetry
$A_{LT'}$ with the experimental data of Ref.\ \cite{Weis:2007kf}
and the results of HBChPT \cite{Bernard:1996ti} and the DMT model \cite{Kamalov:2000en,Kamalov:2001qg}.

   In general, the DMT model gives a very good description of all
observables and amplitudes in the threshold region and can be used
as a guideline for theoretical calculations in cases where
experimental data do not exist.
   The HBChPT calculations shown in Fig.~\ref{fig:weispi0electro} were fitted to these data
and are taken from Ref.~\cite{Weis:2007kf}.
   In contrast, our RChPT calculation is not fitted to these data, as all LECs were already determined
with the data discussed above.
   While HBChPT gives a better description for the unpolarized cross section
$\sigma_0(\Theta_\pi)=\sigma_T(\Theta_\pi)+\epsilon
\sigma_L(\Theta_\pi)$ than our RChPT calculation, a comparison with
the separated cross sections $\sigma_T$ and $\sigma_L$ shows that
this is mainly due to a longitudinal cross section which is much too
small in the HBChPT fit.
   For the other observables $\sigma_{LT}$, $\sigma_{TT}$, and the asymmetry $A_{LT'}$,
RChPT compares much better to the data than HBChPT.
   It is interesting to note that the asymmetry $A_{LT'}$ depends only
weakly on LECs and has an important contribution from the parameter-free pion
loop contribution.

   For the experimental set-up with $\Phi_\pi=90^\circ$, the asymmetry
takes the form [see Eq.~(\ref{eqn:wqparts})]
\begin{equation}
A_{LT'}(\Theta_\pi) =
\frac{\sqrt{2\epsilon(1-\epsilon)}\;\sigma_{LT'}(\Theta_\pi)}{\sigma_{T}(\Theta_\pi)
+\epsilon\; \sigma_{L}(\Theta_\pi)-\epsilon\;
\sigma_{TT}(\Theta_\pi)}\,.
\end{equation}
   Expanding the observables up to and including $P$ waves, we get at
$\Theta_\pi=90^\circ$
\begin{equation}
A_{LT'}(90^\circ) =
\frac{\sqrt{2\epsilon(1-\epsilon)}\;\sqrt{Q^2/k_0^2}\;\mbox{Im}(P_5^*\,E_{0+}
+ L_{0+}^*\,P_2)}{|E_{0+}|^2+\frac{1}{2}(|P_3|^2+|P_2|^2)
+\epsilon\,(Q^2/k_0^2)(|L_{0+}|^2+|P_5|^2)-\epsilon\frac{1}{2}(|P_2|^2-|P_3|^2)}\,,
\end{equation}
where $P_2=3E_{1+}-M_{1+}+M_{1-}$, $P_3=2M_{1+}+M_{1-}$, and
$P_5=L_{1-}-2L_{1+}$.
   As a further simplification, we can assume all
$P$-wave amplitudes as real numbers, where the magnitudes of $P_2$
and $P_3$ are much larger than those of all other multipoles.
   For $\epsilon\approx 1$ we find in very good approximation the simple
form
\begin{equation}
A_{LT'}(90^\circ) \approx
\frac{\sqrt{2\epsilon(1-\epsilon)}\;\sqrt{Q^2/k_0^2}\;(-P_2)\;\mbox{Im}(L_{0+})}{P_3^2}\,.
\end{equation}
   Therefore, this asymmetry is very sensitive to the imaginary part of
the longitudinal $S$ wave $L_{0+}$, hence practically independent
of LECs.
   This is very similar to the case of the target asymmetry $T$ for $\gamma p\to p\pi^0$
which we discussed in our previous article~\cite{Hilt:2013uf}.
   There, the target asymmetry is shown to be the ideal polarization observable
to measure $\mbox{Im}(E_{0+})$.

\subsection{$\gamma+p\rightarrow n+\pi^+$ and $\gamma+n\rightarrow p+\pi^-$}
   We discuss both reaction channels together as we also had to fit them simultaneously.
   In the production of a $\pi^0$ on a proton it is not possible to separate the LECs
into their contributions resulting from isospin (0) and (+) amplitudes.
   In contrast, in the channels involving charged pions one can uniquely determine the LECs in one
of the reaction channels alone, as the kinematic structures of the LECs in the ($-$) component
differ completely from those of the (0) component.
   This is ultimately related to the different crossing behavior of Eq.~(\ref{eqn:crossing})
for the different isospin channels.

   Strictly speaking, the production of a $\pi^-$ on a neutron has never been studied
experimentally as there exists no free neutron target.
   Therefore, one can either study the inverse reaction, namely
radiative pion capture, or use, e.g., deuterium as a target.
   The pion capture cross sections can be completely related to those of pion photoproduction
\cite{Fearing:2000uy}.
   Here, we focus on all existing data of pion production for both channels as no single
experiment contains enough precise data to determine the LECs.
   We take the world data collected in the data base of the SAID program of Ref.\ \cite{SAID}.
   Besides differential cross sections, there also exist data for the photon asymmetry
$\Sigma$, the polarized target asymmetry $T$, and the recoil polarization $P$.
   For the latter two, the existing data points belong to energies which are very high from the
point of view of ChPT.
   Moreover, these two quantities depend strongly on the imaginary
part of the amplitude.
   Therefore, we cannot describe these data points without including
the $\Delta$ resonance and so we do not take them into account.

\begin{figure}[htbp]
    \centering
        \includegraphics[width=\textwidth]{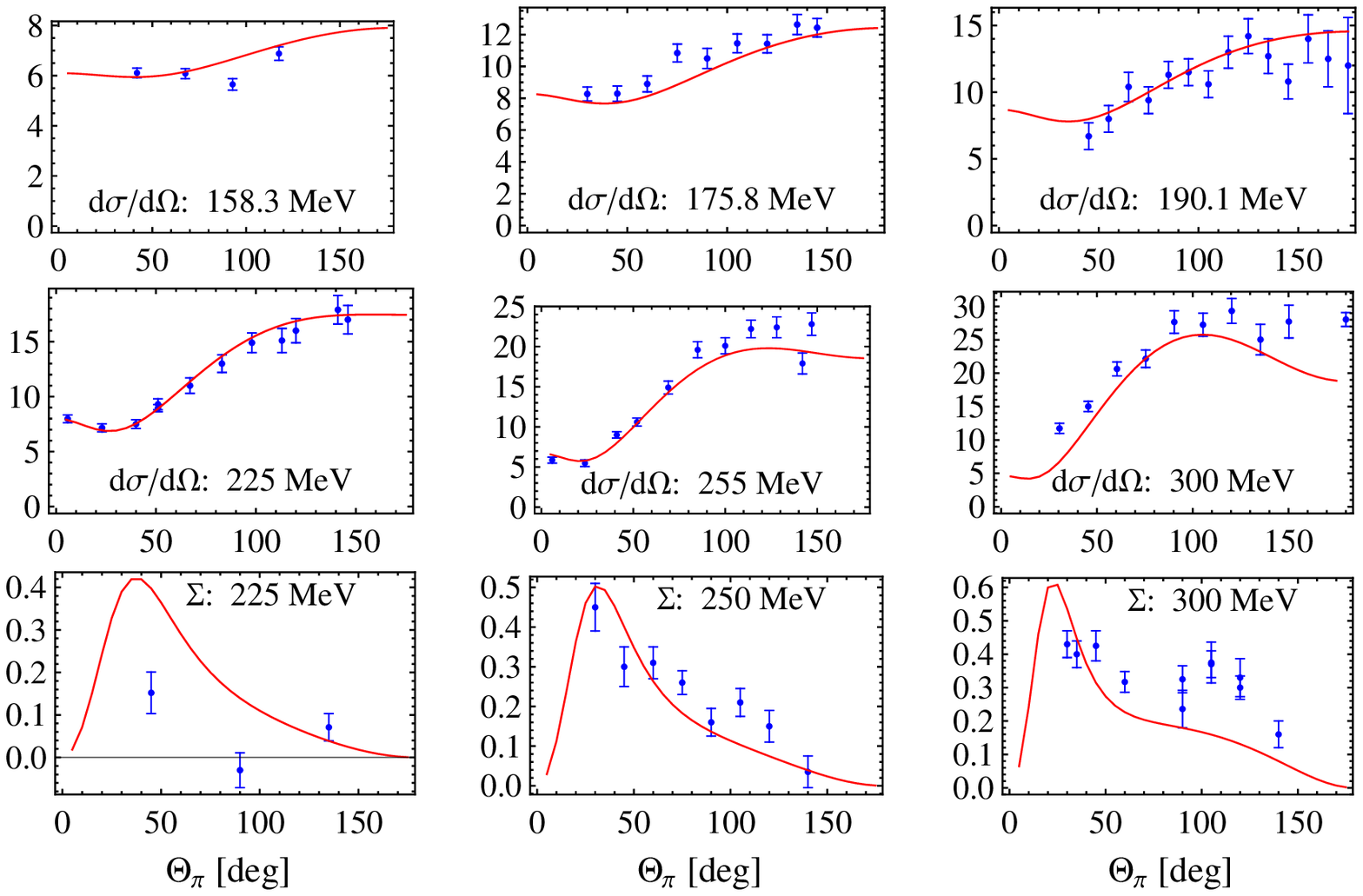}
    \caption{Angular distribution of differential cross sections in $\mu$b/sr and photon asymmetries $\Sigma$
    for $\gamma+n\rightarrow p+\pi^-$ for different $E_\gamma^\textnormal{lab}$.
    The curves show our RChPT calculations at $O(q^4)$.
    The data are from  Refs.\ \cite{SAID,Liu:1964bi,Rossi:1973wf,Kondo:1974yf,Ganenko:1976mf,Fujii:1976jg,
    Salomon:1983xn,Bagheri:1987kf,Wang:1992,Liu:1994}.}
    \label{fig:minusobscharged}
\end{figure}

\begin{figure}[htbp]
    \centering
        \includegraphics[width=\textwidth]{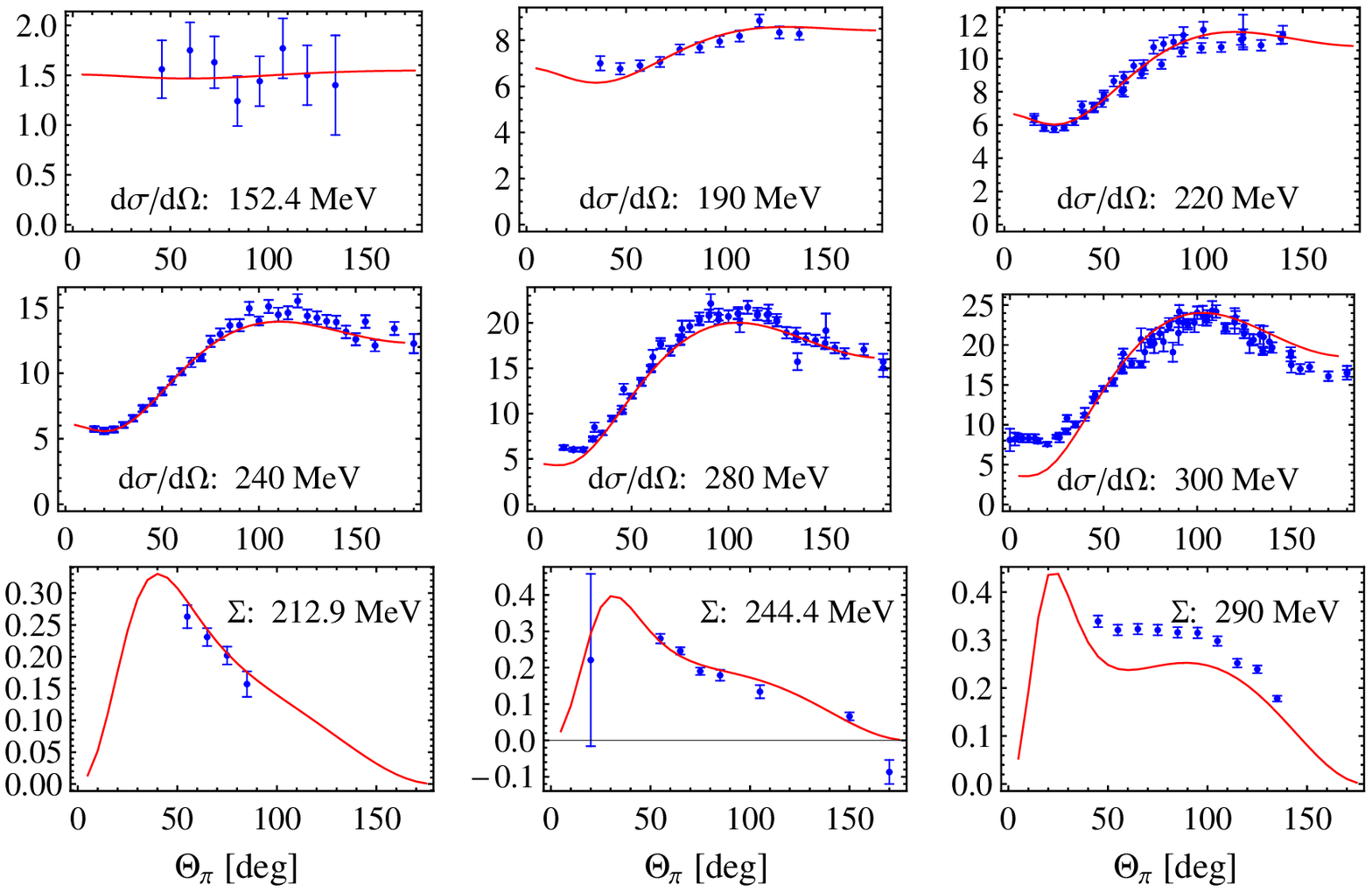}
    \caption{Angular distribution of differential cross sections in $\mu$b/sr and photon asymmetries $\Sigma$
    for $\gamma+p\rightarrow n+\pi^+$ for different $E_\gamma^\textnormal{lab}$.
    The curves show our RChPT calculations at $O(q^4)$.
    The data are from Refs.\ \cite{SAID,Betourne:1968bd,Fischer:1970df,Fischer:1972mt,Fujii:1976jg,Buechler:1994jg,
    Korkmaz:1999sg,Beck:1999ge,Branford:1999cp,Blanpied:2001ae,Ahrens:2004pf}.}
    \label{fig:plusobscharged}
\end{figure}

\begin{figure}[htbp]
    \centering
        \includegraphics[width=0.80\textwidth]{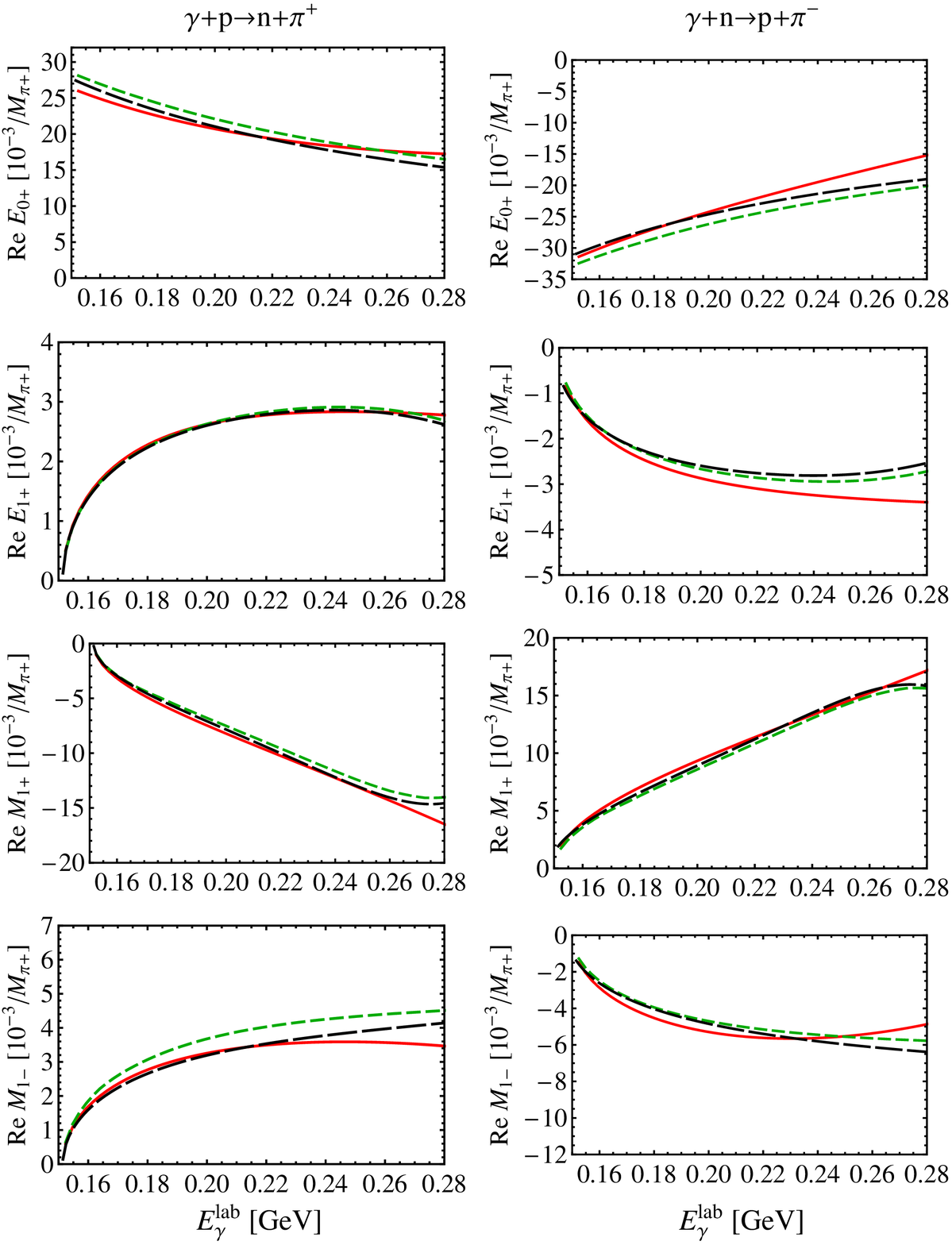}
    \caption{(Color online) $S$- and $P$-wave multipoles for the charged photoproduction channels as a function of
    $E_\gamma^\textnormal{lab}$.
    The solid (red) curves show our RChPT calculations at $O(q^4)$.
    The short-dashed (green) and long-dashed (black) curves are
    the predictions of the DMT model \cite{Kamalov:2000en,Kamalov:2001qg} and the GL model
    \cite{Gasparyan:2010xz}, respectively.}
    \label{fig:spwavesgelphoto}
\end{figure}

   In the charged channels, the $S$ wave dominates in the threshold region
as predicted by the Kroll-Ruderman term \cite{Kroll:1953vq}.
   Above threshold, one needs more partial waves to correctly reproduce the full amplitude, as the pion pole
enhances higher partial waves.
   In our fits we therefore use the full CGLN amplitudes to determine the LECs.
   As before, the procedure relies on multiple fits with random starting values.
   The errors for the fit parameters are again estimated via the bootstrap method.
   We also had to estimate the maximum energy to be used for our fit.
   With the same argument as above, we use $W^\textnormal{max}=1160$ MeV, resulting in $\chi^2_\textnormal{red}=2.39$.
   In Figs.\ \ref{fig:minusobscharged} and \ref{fig:plusobscharged}, we show some exemplary results.
   For the differential cross sections we find a good agreement with the data up to the highest energies we
took into account.
   The asymmetry $\Sigma$ can also be described quite well over the whole energy range.
   Only at energies close to the $\Delta$ resonance deviations become visible.
   Of course, as we have to determine 9 LECs there is some amount of freedom, when fitting the data.
   In order to illustrate that our results are by no means coincidental, in Fig.\ \ref{fig:spwavesgelphoto}
we show the $S$- and $P$-wave multipoles of both channels in comparison with the
DMT model \cite{Kamalov:2000en} and the covariant, unitary, chiral approach of GL \cite{Gasparyan:2010xz}.
   One can clearly see in the $E_{1+}$ and $M_{1+}$ multipole that we did not
include the $\Delta$ resonance explicitly, because the real parts of the multipoles should have a zero crossing
at the $\Delta$ resonance position $E_\gamma^\textnormal{lab}=0.34$ GeV, which is indicated in
GL and DMT, as both multipoles start to drop off at the highest energies shown here.
   The small discrepancies between our calculation and the other two models can be traced back to
the data we used.
   In order to determine $E_{0+}$ and the $P$ waves correctly, one not only needs precise data for
the differential cross sections but also for the asymmetry.
   Here, we only have few data for the asymmetry available and, furthermore, their relative error
is bigger compared to the cross sections, which lowers their weight in the fit.

\subsection{$\gamma^{(*)}+p\rightarrow n+\pi^+$}
\begin{figure}[htbp]
    \centering
        \includegraphics[width=\textwidth]{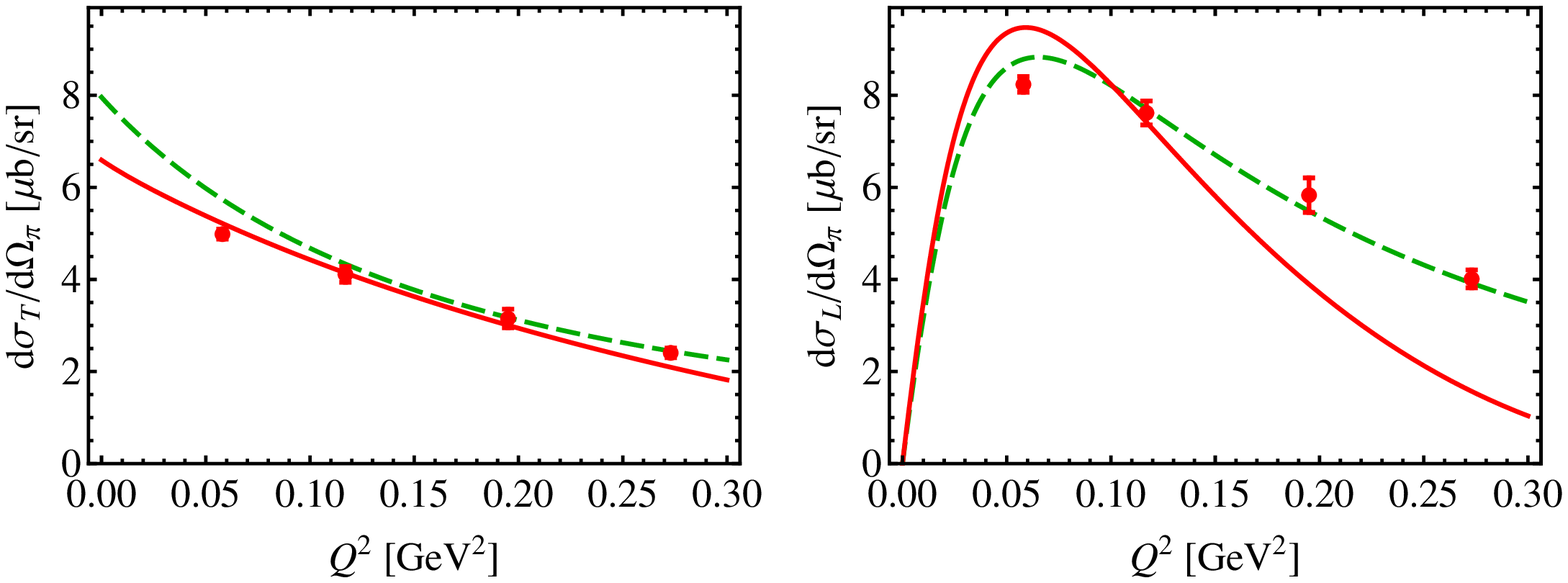}
    \caption{(Color online) Differential cross sections as a function of $Q^2$ for $\gamma^*+p\rightarrow n+\pi^+$ at
    $W=1125$ MeV and $\Theta_\pi=0^\circ$.
    The solid (red) curves show our RChPT calculation at $O(q^4)$ and the dashed (green)
    curves are the predictions of the DMT model \cite{Kamalov:2000en,Kamalov:2001qg}.
    The data are from Ref.\ \cite{baumann,Drechsel:2007sq}.}
    \label{fig:chelectro}
\end{figure}

   In this reaction channel, only a few data points exist in the energy
range and for photon virtualities, where ChPT can be applied.
   Unfortunately, these data of the differential cross  sections
$\sigma_T$ and $\sigma_L$ at $W=1125$ MeV are at one fixed angle,
namely $\Theta_\pi=0^\circ$ \cite{baumann,Drechsel:2007sq}.
   Hence, the angular distribution cannot be analyzed.
   Nevertheless, we use the forward-scattering cross section to
fix the two remaining constants through a fit to the data.
   We refrain from giving an error for these LECs as this is only a first
estimate and the amount of data is too small for good statistics.
   The results of our calculation are shown in Fig.\ \ref{fig:chelectro}.
   While the theory agrees with the data for $\sigma_T$, for $\sigma_L$ some
deviation is visible.

\begin{table}[htbp]
\centering
\caption{Numerical values of all LECs of pion photo- and electroproduction.
The $\ast$ indicates constants that appear in electroproduction, only.
If possible, the errors were estimated using the bootstrap method (see text for details).
In the case of the electroproduction LECs $e_{52}$, $e_{53}$, $e_{72}$, and $e_{73}$ we can only give
errors for $\tilde{e}_{52}$ and $\tilde{e}_{53}$
(compare Table \ref{tab:lecvaluespi0electro} and text).}
 \label{tab:lecvalues}
        \begin{tabular}{ccc}
\hline
\hline
  Isospin channel   & LEC  &      Value       \\
\hline
 0 & $d_9\ [\textnormal{GeV}^{-2}]$               &       $-1.22\pm0.12$            \\
 0 & $e_{48}\ [\textnormal{GeV}^{-3}]$            &       $5.2\pm1.4$            \\
 0 & $e_{49}\ [\textnormal{GeV}^{-3}]$            &       $0.9\pm2.6$            \\
 0 & $e_{50}\ [\textnormal{GeV}^{-3}]$            &       $2.2\pm0.8$            \\
 0 & $e_{51}\ [\textnormal{GeV}^{-3}]$            &       $6.6\pm3.6$            \\
 0 & $e_{52}^*\ [\textnormal{GeV}^{-3}]$          &       $-4.1$            \\
 0 & $e_{53}^*\ [\textnormal{GeV}^{-3}]$          &       $-2.7$            \\
 0 & $e_{112}\ [\textnormal{GeV}^{-3}]$           &       $9.3\pm1.6$            \\
\hline
 + & $d_8\ [\textnormal{GeV}^{-2}]$               &       $-1.09\pm0.12$            \\
 + & $e_{67}\ [\textnormal{GeV}^{-3}]$            &       $-8.3\pm1.5$            \\
 + & $e_{68}\ [\textnormal{GeV}^{-3}]$            &       $-0.9\pm2.6$            \\
 + & $e_{69}\ [\textnormal{GeV}^{-3}]$            &       $-1.0\pm2.2$            \\
 + & $e_{71}\ [\textnormal{GeV}^{-3}]$            &       $-4.4\pm3.7$            \\
 + & $e_{72}^*\ [\textnormal{GeV}^{-3}]$          &       $10.5$            \\
 + & $e_{73}^*\ [\textnormal{GeV}^{-3}]$          &       $2.1$            \\
 + & $e_{113}\ [\textnormal{GeV}^{-3}]$           &       $-13.7\pm2.6$            \\
\hline
 $-$ & $d_{20}\ [\textnormal{GeV}^{-2}]$            &       $4.34\pm0.08$            \\
 $-$ & $d_{21}\ [\textnormal{GeV}^{-2}]$            &       $-3.1\pm0.1$            \\
 $-$ & $e_{70}\ [\textnormal{GeV}^{-3}]$            &       $3.9\pm0.3$            \\
\hline
\hline
\end{tabular}

\end{table}

\section{Chiral MAID}
   The complete amplitude at $O(q^4)$ is a rather lengthy expression and cumbersome to handle.
   Nevertheless, we wanted to give easy access to anybody who is interested in this calculation.
   We therefore created $\chi$MAID which is a web interface with certain underlying FORTRAN programs.
   The main parts of these programs are adopted from MAID2007 \cite{Drechsel:2007if}.
   Unfortunately, the computing time for the desired quantities, e.g., cross sections or multipoles,
is too high for a web-based application.
   We avoid calculating the complete amplitude by restricting the input for $\chi$MAID to multipoles
up to and including $l=4$ or, in other words, $G$ waves.
   All observables are derived from the multipoles which we computed beforehand for all reaction channels.
   The multipoles are calculated for an energy range of $W=1073.3-1190$ MeV and, for
electroproduction, through $Q^2=0.3\ \textnormal{GeV}^2$.

   The loop contributions, including their parameters, are fixed and cannot be modified from the outside.
   On the other hand, the contact diagrams at ${O}(q^3)$ and ${O}(q^4)$ enter analytically and the
corresponding LECs can be changed arbitrarily (see Table \ref{tab:lecvalues} for our present values).
   This is an important feature in the light of future new experimental data, as more precise data help
to get better access to the LECs (see Appendix \ref{treeresults}).

   Of course,  $\chi$MAID has a limited range of applicability.
   First of all, ChPT without additional dynamical degrees of freedom restricts the
energy region, where our results can be applied.
   In the case of neutral pion photoproduction (see Section \ref{pi0photo}) one can clearly see that for energies
above $E_\gamma^\textnormal{lab}\approx 170$ MeV the theory starts to deviate
from experimental data.
   In the case of the charged channels the range of applicability is larger, but some observables are
quite sensitive to the cutoff of multipoles, as the pion pole term is important at small angles.
   As an estimate, for $W>1160$ MeV the difference between our full amplitude and the approximation
up to and including $G$ waves becomes visible.

\section{Summary}
   We have presented and discussed a full $O(q^4)$ calculation of pion photo- and electroproduction
in the framework of manifestly Lorentz-invariant (relativistic) chiral perturbation theory.
   By performing fits to the available experimental data, we determined all 19 LECs of the contact graphs
at $O(q^3)$ and $O(q^4)$ (see Table \ref{tab:lecvalues}).
   Our findings can be summarized as follows.

   The latest data of Ref.\ \cite{Hornidge:2012ca} for $\pi^0$
photoproduction on a proton gave us the possibility to determine the
$S$ and $P$ waves in the threshold region.
   The first measurement of the photon asymmetry $\Sigma$ starting from threshold was an
important feature of this experiment, because only this way one can
access the $S$ and $P$ waves simultaneously.
   In principle, a sixth LEC exists at $O(q^4)$ which mainly affects the $E_{2-}$ multipole.
   Unfortunately, we were not able to pin down this constant and, therefore, neglected it in our
fit.
   Nevertheless, we found a good agreement with the observables and the multipoles up to
$E_\gamma^\textnormal{lab}\approx 170$ MeV.

   The experiment of Ref.\ \cite{Merkel:2011cf} was utilized to determine the two remaining LECs
for $\pi^0$ electroproduction on the proton.
   We found that our results are in good agreement with the available data including the total cross sections.
   It will be interesting to compare our results with future experiments when further observables can
be measured.

   In the case of charged pion photoproduction we had to examine both
reaction channels simultaneously.
   None of the existing experiments covered a large enough energy range and, therefore, we
decided to use the global data available \cite{SAID}.
   The description of the differential cross sections turned out to be satisfactory almost
up to the $\Delta$ resonance region.
   This finding is a little bit misleading as one can clearly see from the results
for the asymmetry $\Sigma$, where we find that deviations occur already at
somewhat lower energies.
   Furthermore, the large number of LECs subsume some of
the missing imaginary part of the amplitude.
   This can lead to a wrong picture at the highest energies we took into account,
as the missing piece of the imaginary part of the amplitude
cannot be included in the LECs.

   For charged pion electroproduction we found only one experiment which was suited
for an analysis in terms of ChPT \cite{baumann,Drechsel:2007sq}.
   There are only few data points available which, in addition, were
measured at a fixed angle, namely in the forward direction.
   Hence, we consider our analysis only a first estimate.
   Future experiments will hopefully give us the opportunity to reexamine the two LECs
remaining for charged pion electroproduction.

   Finally, we presented the web interface $\chi$MAID \cite{website}.
   With the estimates for the LECs presented in this article, one can obtain predictions
for all desired observables in the threshold region.
   Therefore, we refrained from showing any additional predictions here.
   It is clear that new experiments will lead to different estimates for the LECs.
   For that reason, we included in  $\chi$MAID the possibility to change the LECs arbitrarily.
   This will help to further study the range of validity and applicability of ChPT in the future.

\begin{acknowledgements}
   This work was supported by the Deutsche Forschungsgemeinschaft (SFB 443 and 1044).
   The authors would like to thank D.~Drechsel, J.~Gegelia, and D.~Djukanovic for useful
discussions and support.
   We also thank the A2 and CB-TAPS collaborations for making available the experimental
data prior to publication.
\end{acknowledgements}

\begin{appendix}
\section{Connection between the different sets of amplitudes of pion electroproduction\label{ampcon}}
   The connection between the Ball amplitudes $B_i$ of Eq.~(\ref{eqn:ball}) and the invariant amplitudes
$A_i$ of Eq.~(\ref{eqn:dennery}) can be derived by equating the two parameterizations:
\begin{equation}
\sum_{i=1}^6 A_i M_i^\mu\stackrel{!}{=}\sum_{i=1}^8 B_i V_i^\mu.
\end{equation}
   In order to connect the two sets, one can make use of current conservation to eliminate two of the Ball
amplitudes [see Eqs.\ (\ref{eqn:stromerhalt2})].
   By comparing the Lorentz structures, one can read off the following connection
[here, we replaced $B_1$ and $B_2$ by exploiting Eqs.\ (\ref{eqn:stromerhalt2})]:
\begin{equation}
\begin{split}
A_1&=i(B_5+m_N B_6),\\
A_2&=-i\frac{k\cdot qB_3+k^2(B_4+B_5)}{k\cdot P(k^2-2k\cdot q)},\\
A_3&=iB_7,\\
A_4&=\frac{i}{2}B_6,\\
A_5&=\frac{i}{k^2-2k\cdot q}(B_3+2B_4+2B_5),\\
A_6&=-iB_8.
\end{split}
\end{equation}
   In a similar manner one can relate the invariant amplitudes $A_i$ of Eq.~(\ref{eqn:dennery})
to the CGLN amplitudes ${\cal F}_i$ of Eq.~(\ref{eq:F}).
   In the following equations, all (non-invariant) quantities are defined in the cm frame:
\begin{equation}
\begin{split}
\mathcal{F}_1&=\frac{W-m_N}{8\pi W}\sqrt{(E_i+m_N)(E_f+m_N)}\Big[A_1+(W-m_N)A_4\\
&\quad-\frac{2m_N \nu_B}{W-m_N}(A_3-A_4)+\frac{Q^2}{W-m_N}A_6\Big],\\
\mathcal{F}_2&=\frac{W+m_N}{8\pi W}|\vec{q}\hspace{0.1cm}|\sqrt{\frac{E_i-m_N}{E_f+m_N}}\Big[-A_1+(W+m_N)A_4\\
&\quad-\frac{2m_N \nu_B}{W+m_N}(A_3-A_4)+\frac{Q^2}{W+m_N}A_6\Big],\\
\mathcal{F}_3&=\frac{W+m_N}{8\pi W}|\vec{q}\hspace{0.1cm}
|\sqrt{(E_i-m_N)(E_f+m_N)}\left[\frac{2W^2-2m_N^2+Q^2}{2(W+m_N)}A_2\right.\\
&\quad\left.+A_3-A_4-\frac{Q^2}{W+m_N}A_5\right],\\
\mathcal{F}_4&=\frac{W-m_N}{8\pi W}|\vec{q}\hspace{0.1cm}
|^2\sqrt{\frac{E_i+m_N}{E_f+m_N}}\left[-\frac{2W^2-2m_N^2+Q^2}{2(W-m_N)}A_2\right.\\
&\quad\left.+A_3-A_4+\frac{Q^2}{W-m_N}A_5\right],\\
\mathcal{F}_5&=\frac{k_0}{8\pi W}\sqrt{\frac{E_f+m_N}{E_i+m_N}}\Big\{(E_i+m_N)A_1\\
&\quad+\left[4m_N \nu_B\left(W-\frac{3}{4}k_0\right)
-\vec{k}^2W+E_\pi\left(W^2-m_N^2+\frac{1}{2}Q^2\right)\right]A_2\\
&\quad\left.+[E_\pi(W+m_N)+2m_N\nu_B]A_3\right.\\
&\quad+[(E_i+m_N)(W-m_N)-E_\pi(W+m_N)-2m_N\nu_B]A_4\\
&\quad+(2m_N \nu_Bk_0-E_\pi Q^2)A_5\\
&\quad-(E_i+m_N)(W-m_N)A_6\Big\},\\
\mathcal{F}_6&=\frac{k_0|\vec{q}\hspace{0.1cm}|}{8\pi W\sqrt{(E_f+m_N)(E_i-m_N)}}\Big\{-(E_i-m_N)A_1\\
&\quad+\left[\vec{k}^2W-4m_N\nu_B\left(W-\frac{3}{4}k_0\right)
-E_\pi\left(W^2-m_N^2+\frac{1}{2}Q^2\right)\right]A_2\\
&\quad+[E_\pi(W-m_N)+2m_N\nu_B]A_3\\
&\quad+[(E_i-m_N)(W+m_N)-E_\pi(W-m_N)-2m_N\nu_B]A_4\\
&\quad+(E_\pi Q^2-2m_N\nu_Bk_0)A_5\\
&\quad-(E_i-m_N)(W+m_N)A_6\Big\},
\end{split}
\end{equation}
where $2\nu_B=-k\cdot q$.

\section{differential cross sections\label{diffcs}}
   In order to describe the individual parts of the differential cross section, the so-called
response functions are commonly used \cite{Drechsel:1992pn}:
\begin{equation}
\begin{split}
R_T&=|F_1|^2+|F_2|^2+\frac{1}{2}\sin^2{\Theta_\pi}\left(|F_3|^2+|F_4|^2\right)\\
&\quad-\textnormal{Re}\left[2\cos\Theta_\pi F_1^*F_2-\sin^2\Theta_\pi(F_1^*F_4+F_2^*F_3+\cos\Theta_\pi F_3^*F_4)\right],\\
R_L&=|F_5|^2+|F_6|^2+2\cos\Theta_\pi\textnormal{Re}\left(F_5^*F_6\right),\\
R_{LT}&=-\sin\Theta_\pi\textnormal{Re}\left[(F_2^*+F_3^*+\cos\Theta_\pi F_4^*)F_5
+(F_1^*+F_4^*+\cos\Theta_\pi F_3^*)F_6\right],\\
R_{TT}&=\sin^2\Theta_\pi\left[\frac{1}{2}(|F_3|^2+|F_4|^2)
+\textnormal{Re}\left(F_1^*F_4+F_2^*F_3+\cos\Theta_\pi F_3^*F_4\right)\right],\\
R_{LT'}&=-\sin\Theta_\pi\textnormal{Im}\left[(F_2^*+F_3^*+\cos\Theta_\pi F_4^*)F_5
+(F_1^*+F_4^*+\cos\Theta_\pi F_3^*)F_6\right].
\end{split}
\end{equation}
   All other parts of the differential cross section are not relevant for the discussions in this article.
   They can be found in Ref.\ \cite{Drechsel:1992pn}.
   The connection between the response functions and the cross sections reads
\begin{equation}
\begin{split}
\frac{d\sigma_T}{d\Omega_\pi}&=\frac{|\vec{q}|}{k_\gamma^\textnormal{cm}}R_T,\\
\frac{d\sigma_L}{d\Omega_\pi}&=\frac{|\vec{q}|}{k_\gamma^\textnormal{cm}}\frac{Q^2}{k_0^2}R_L,\\
\frac{d\sigma_{LT}}{d\Omega_\pi}&=\frac{|\vec{q}|}{k_\gamma^\textnormal{cm}}\frac{Q}{|k_0|}R_{LT},\\
\frac{d\sigma_{TT}}{d\Omega_\pi}&=\frac{|\vec{q}|}{k_\gamma^\textnormal{cm}}R_{TT},\\
\frac{d\sigma_{LT'}}{d\Omega_\pi}&=\frac{|\vec{q}|}{k_\gamma^\textnormal{cm}}\frac{Q}{|k_0|}R_{LT'},
\end{split}
\end{equation}
where $k_\gamma^\textnormal{cm}=k_\gamma m_N/W$ is the photon equivalent energy in the cm frame.
   Furthermore, several polarization observables can be derived.
   Here, we only need the photon asymmetry,
\begin{equation}
\Sigma=-R_{TT}/R_T.
\end{equation}
   It appears in the case of polarized photons, as the differential cross section
$d\sigma/d\Omega_\pi$ in the cm frame then gets modulated depending on the angle $\phi$
between the polarization vector of the photon and the reaction plane spanned by the nucleon and
pion three momenta:
\begin{equation}
\frac{d\sigma}{d\Omega_\pi}(\Theta_\pi,\phi)=\frac{d\sigma}{d\Omega_\pi}(\Theta_\pi)
\left[1-\Sigma(\Theta_\pi)\cos(2\phi)\right].
\end{equation}

\section{Making new estimates for LECs \label{treeresults}}
   If one is interested in analyzing new experiments and re-estimating the LECs,
one can proceed as follows.
   By switching off the LECs on the $\chi$MAID web page, one gets any desired amplitude or the
multipoles with $l\leq 2$ numerically.
   One can then add the analytic expressions for the contact diagrams given below.
   From that one can calculate any desired observable and make estimates for the LECs.
   We give the results for the invariant amplitudes, as they are in a very compact form.
   The results for the $O(q^3)$ contact diagrams read
\begin{equation}
\begin{split}
A_1^{(0)}&=-\frac{2 e  d_9 t}{F m},\\
A_2^{(0)}&=-\frac{2 e d_9 \left(k^2-M^2+t\right)}{F m\left(M^2-t\right)},\\
A_3^{(0)}&=0,\\
A_4^{(0)}&=-\frac{4 e d_9}{F},\\
A_5^{(0)}&=\frac{e d_9(s-u)}{F m\left(M^2-t\right)},\\
A_6^{(0)}&=0,\\
A_1^{(-)}&=\frac{e d_{20}  (s-u)}{4 F m},\\
A_2^{(-)}&=0,\\
A_3^{(-)}&=-\frac{e d_{20} \left(k^2-M^2-t\right)}{8Fm^2}+\frac{ed_{21}}{F},\\
A_4^{(-)}&=\frac{e d_{20} (s-u) }{8 F m^2},\\
A_5^{(-)}&=0,\\
A_6^{(-)}&=-\frac{e d_{20}\left(k^2-M^2-t\right)}{8 F m^2}.
\end{split}
\end{equation}
The results for the (+) components can be derived from the (0) components by replacing $d_9\rightarrow d_8$.
   From a practical point of view, one may replace $m$, $F$, and $M^2$ by their physical values
$m_N$, $F_\pi$, and $M_\pi^2$, as the consequences for the pion production amplitude are of higher order
in the chiral expansion.
   For the $O(q^4)$ contact diagrams the results read
\begin{equation}
\begin{split}
A_1^{(0)}&=-e(2e_{49}-e_{51})\frac{k^2+M^2-t}{2F}\\
&\quad-\frac{e e_{50}}{6F m^2}\left\{2 M^4+\left[4m^2-2(s+u)\right] M^2+2m^4+s^2+u^2-2m^2 (s+u)\right\}\\
&\quad+e e_{52}\frac{k^2}{F}-2ee_{112}\frac{M^2}{F},\\
A_2^{(0)}&=-\frac{2 e e_{49} \left(k^2+M^2-t\right)}{F\left(M^2-t\right)}
+\frac{2e e_{52}k^2}{F\left(M^2-t\right)},\\
A_3^{(0)}&=e(e_{48}+e_{49})\frac{s-u}{2 F m}
+e e_{50}\frac{(k^2-M^2-t)(s-u)}{12 F m^3},\\
A_4^{(0)}&=-e(2e_{49}-e_{51})\frac{k^2+M^2-t}{4Fm}\\
&\quad-\frac{e e_{50}}{12F m^3}\left\{2 M^4+\left[4m^2-2(s+u)\right] M^2+2m^4+s^2+u^2-2m^2 (s+u)\right\}\\
&\quad+e e_{52}\frac{k^2}{2Fm}-ee_{112}\frac{M^2}{Fm},\\
A_5^{(0)}&=\frac{e \left(e_{49}-e_{52}\right)(s-u)}{F\left(M^2-t\right)},\\
A_6^{(0)}&=\frac{e e_{50}\left(k^2-M^2-t\right)(s-u)}{12 F m^3}
+\frac{e \left(e_{52}+e_{53}\right) (s-u)}{4Fm},\\
A_1^{(-)}&=-\frac{e e_{70} t (s-u)}{F m^2},\\
A_2^{(-)}&=-\frac{e e_{70}\left(k^2-M^2+t\right)(s-u)}{F m^2 \left(M^2-t\right)},\\
A_3^{(-)}&=0,\\
A_4^{(-)}&=-\frac{2 e e_{70} (s-u)}{F m},\\
A_5^{(-)}&=\frac{e e_{70} (s-u)^2}{2 F m^2\left(M^2-t\right)},\\
A_6^{(-)}&=0.
\end{split}
\end{equation}
Again, the expressions for the (+) components follow from the (0) components by making the following replacements:
\begin{align*}
e_{48}&\rightarrow e_{67},\\
e_{49}&\rightarrow e_{68},\\
e_{50}&\rightarrow e_{69},\\
e_{51}&\rightarrow e_{71},\\
e_{52}&\rightarrow e_{72},\\
e_{53}&\rightarrow e_{73},\\
e_{112}&\rightarrow e_{113}.
\end{align*}
   Moreover, $m$, $F$, and $M^2$ may be replaced by their physical values.
\end{appendix}

\end{document}